\documentclass[a4paper,fleqn,usenatbib]{mnras}

\usepackage[T1]{fontenc}
\usepackage{ae,aecompl}

\usepackage{graphicx}	% Including figure files
\usepackage{multirow}
\usepackage{txfonts}
\usepackage{float}
\def\aj{AJ}
\def\apj{ApJ}
\def\aap{A\&A}
\def\apjl{ApJL}
\def\apjs{ApJS}

\def\mnras{MNRAS}

\title[Merger traces in the Fornax dSph galaxy]{Merger traces in the spatial distribution of stellar populations in the Fornax dSph galaxy}

\author[Andr\'es del Pino et al.]{Andr\'es del Pino$^{1,2}$\thanks{E-mail: \mbox{adpm@camk.edu.pl} (ADPM); \mbox{aaj@iac.es} (AA); \mbox{shidalgo@iac.es} (SH).}, Antonio Aparicio$^{2,3}$\footnotemark[1], Sebastian L. Hidalgo$^{2,3}$\footnotemark[1].\\ 
$^1$ Nicolaus Copernicus Astronomical Centre of the Polish Academy of Sciences. ul. Bartycka 18 00-716, Warsaw.\\
$^2$ Instituto de Astrof\'\i sica de Canarias. Calle V\'\i a L\'actea s/n. E38200 - La Laguna, Tenerife, Canary Islands, Spain.\\
$^3$ University of La Laguna. Avda. Astrof\'isico Fco. S\'anchez, s/n. E38206, La Laguna, Tenerife, Canary Islands, Spain.}

\date{Accepted 2015 September 17. Received 2015 September 16; in original form 2015 July 8}

\pubyear{2015}

\begin{document}
\label{firstpage}
\pagerange{\pageref{firstpage}--\pageref{lastpage}}
\maketitle

\begin{abstract}
We present a comprehensive and detailed study of the stellar populations of the Fornax dwarf spheroidal galaxy. We analyse their spatial distributions along the main body of the galaxy, obtaining their surface density maps, together with their radial density profiles. Results are based on the largest and most complete catalogue of stars in Fornax, with more than $3.5\times10^{5}$ stars covering the main body of the galaxy up to $V \sim 24$. We find a differentiated structure in Fornax depending on the stellar ages. Old stars ($\gtrsim 10$ Gyr) follow an elliptical distribution well fitted by King profiles with relatively large core radius ($r_{\rm c} = 760\pm60$ pc). On another hand, young populations ($\lesssim 3$ Gyr) concentrate in the central region of the galaxy ($r_{\rm c} = 210\pm10$ pc), and are better fitted by S\'ersic profiles with $0.8 < n < 1.2$, indicating some discy shape. These stars show strong asymmetries and substructures not aligned with the main optical axes of Fornax. This together with the observed differences between metallicity and age distribution maps strongly suggests accretion of material with different angular momentum. These results lead us to propose a scenario in which Fornax has suffered a major merger at $z\sim1$.
\end{abstract}


%%%%%%%%%%%%%%%%%%%%%%%%%%%%%%%%%%%%%%%%%%%%%%%%%%

%%%%%%%%%%%%%%%%% BODY OF PAPER %%%%%%%%%%%%%%%%%%

\section{Introduction}\label{introduccion}

How galaxies were formed and evolved is still an unsolved question. The introduction of the $\Lambda$-cold dark matter ($\Lambda$CDM) cosmological scenario has stressed the attention on the smallest of these systems: the dwarf galaxies. From a cosmological point of view, dwarfs would be the first galaxies ever formed, from which bigger structures would have formed via mergers \citep*{Blumenthal1985, Dekel1986, Navarro1995, Moore1998}. Since they were the smallest galaxies known until the discovery of the ultra faint galaxies, the dwarf spheroidal galaxies (dSph) were proposed as the \textit{building blocks} of the bigger systems.\\

DSphs are systems characterized by low surface luminosities ($\rm \sum_V \la 0.002~L_\odot~pc^{-2}$), small sizes (about a few hundred of parsecs), and lack of gas. The relatively large velocity dispersions observed in the dSphs, exceeding 7 $\rm km~s^{-1}$ \citep*[see][and references therein]{Aaronson1983, Mateo1998}, suggest the presence of abundant dark matter in them. Assuming to be virialized systems, dSphs reach mass-to-light ratios of $\sim 5 -500$ in solar units \citep{Kleyna2001, Kleyna2005, Odenkirchen2001}. All these discoveries have given rise to competing interpretations concerning their origin and cosmological significance.\\

DSphs had  been traditionally considered as rather simple and evolved systems, dynamically supported by the random motion of their stars (pressure supported). Their spheroidal shape, together with the apparent lack of gas, has contributed to this idea. Notwithstanding, each dSph appears to be unique in its star formation history (SFH), and chemical evolution \citep[see for example][]{Weisz2014}. In addition rotation signatures have been reported in some dSph galaxies \citep[see][]{Battaglia2008b, Fraternali2009, Ho2012}. As a whole, dSphs show a panoply of properties which are still under debate.\\

Several mechanisms have been proposed to be affecting the formation and evolution of these systems. The most relevant include local processes, such as supernova (SN) feedback and interactions with nearby systems, as well as global cosmic environmental factors like the early reionization of the Universe by UV radiation \citep*{Hayashi2003, Taffoni2003, Kravtsov2004, Kazantzidis2011}. Different combinations of these together with other mechanisms may be implied in the evolution of dSph galaxies, causing the large variety in their observed properties.\\

In the present work, we revisit the Fornax dSph, one of the nine classic dSph satellites of the Milky Way (MW). Fornax is, after the Sagittarius dSph, the largest and most luminous of the dSphs MW companions, with a core radius of $\sim$ 550 pc. It is located at a distance of 136$\pm$5 kpc \citep*{Mackey2003, Greco2007, Greco2009, Tammann2008, Poretti2008}, and its principal baryonic component is stellar, with an ambiguous detection of $\rm H_I$ \citep*{Bouchard2006} possibly associated with the MW. The dynamical mass within the observed half-light radius has been estimated to be $\rm 5.6\times10^7 M_\odot$ \citep[][and references therein]{McConnachie2012}.\\

Fornax is a remarkable object to study. It hosts five globular clusters and shows two conspicuous star clumps located at $17^\prime$ and at 1.3$^{\circ}$ from its centre. How these clumps were formed is subject of debate. Some authors claim that these structures are the result of a past merger \citep{Coleman2004, Coleman2005, Coleman2008, Amorisco2012}. On another hand, \citet{deBoer2013} claim that these clumps are more likely to be the result of the quiet infall of gas previously expelled by Fornax during its star formation episodes. Its SFH is complex and long-standing \citep{delPino2013}, while its dynamics has turned out as rather complex \citep{Walker2009, Amorisco2012}.\\

In the present work we carry out a comprehensive and deep study of the morphological properties of the Fornax dSph galaxy. We trace the spatial distribution of different stellar populations, and study metallicity and velocity gradients along the main body of the galaxy. In Section~\ref{Cap:Data}, the data set is presented. Section~\ref{Cap:CMD} describes the colour magnitude diagram (CMD), together with the method followed for sampling the different stellar populations. In section~\ref{Cap:2DMaps}, the two-dimensional distribution of the stellar populations are shown. Section~\ref{Cap:Radial_Profiles} explains the procedure for obtaining the density radial profiles. Besides this, results from isopleth and model fitting are discussed. In Section~\ref{Cap:Interaction_Discussion}, the results are discussed and several scenarios of interaction of Fornax with other systems are proposed. Finally, a summary and the main conclusions of the work are presented in Section~\ref{Cap:Summary}.\\

\section{The data}\label{Cap:Data}

\begin{figure}
\begin{center}
\includegraphics[scale=0.75]{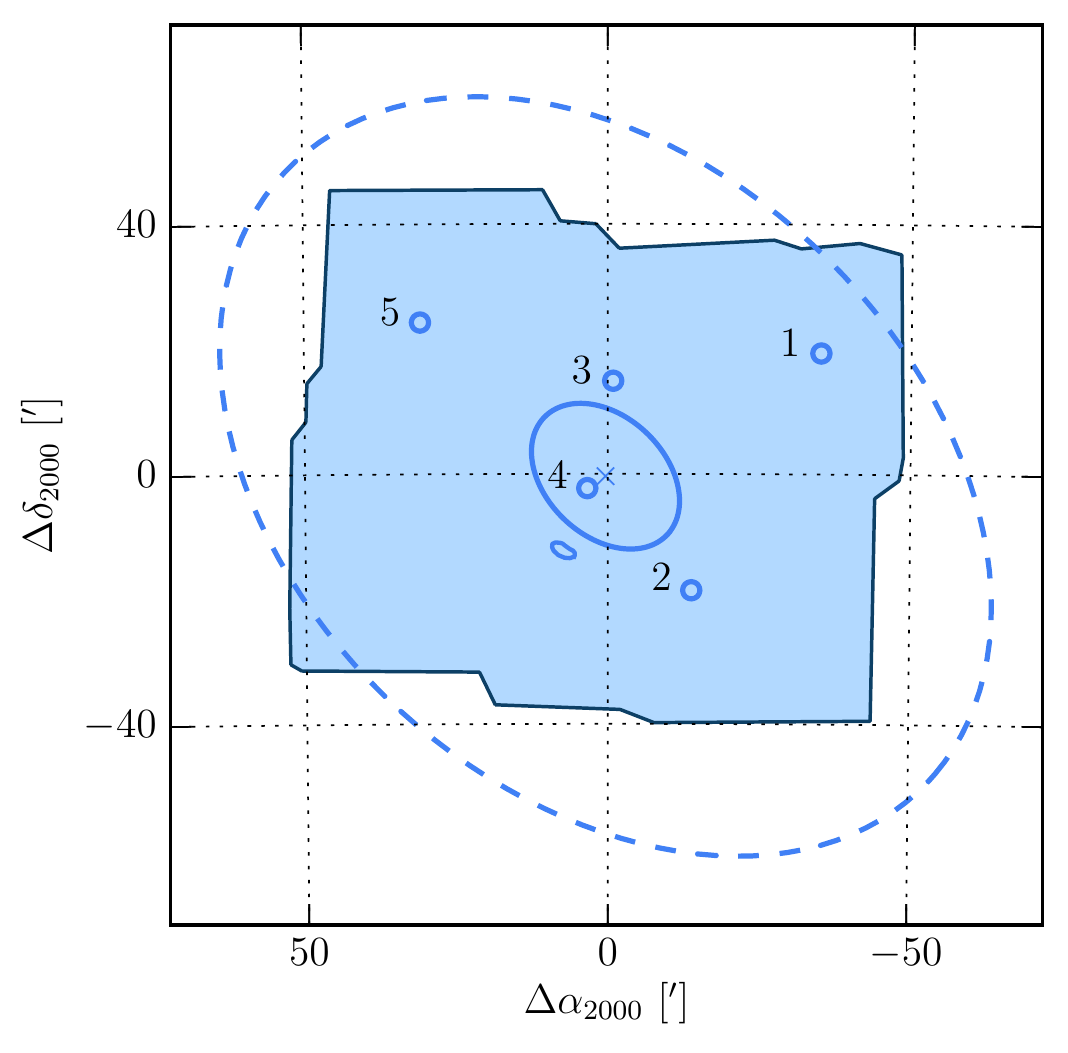}
\caption[A schematic view of the wide field photometry data]{A schematic view of the spatial coverage of the data sets. The blue shaded area corresponds to the photometry sky coverage. Blue dashed-lined ellipse represent the tidal radius of Fornax ($71.4^\prime$), while its core radius ($13.8^\prime$) is shown by a continuous-lined ellipse, both with a position angle of 41$^{\circ}$ \citep{Mateo1998}. The centre of Fornax derived by \citet{van_den_Bergh1999} is represented by a blue cross ($\alpha_{2000} =$ 2h 39$^\prime$ 53.1$^{\prime\prime}$, $\delta_{2000} =$ -34$^{\circ}$ 30$^\prime$ 16.0$^{\prime\prime}$). Small blue open circles mark the position of globular clusters and are labelled accordingly. The position of the shell found by \citet{Coleman2004} is marked at the south-east of the core of the galaxy.}
\label{fig:Data}
\end{center}
\end{figure}

There exist a large amount of photometric data for Fornax, collected with a large variety of scientific objectives, field of view, and data depth, the two latter running normally in opposite directions. With the aim of covering the largest possible area of the galaxy, we have compiled the biggest existing photometric catalogue for this galaxy, with more than $3.5\times10^{5}$ stars. This photometric list is based on the calibrated photometry obtained by \citet{Stetson2000, Stetson2005} and \citet{deBoer2012}. The first catalogue (Cat1) consists in an extensive ensemble of archive data, and comprises about $3.5\times10^{5}$ stars. The second one (Cat2) contains photometry from MOSAIC@CTIO for about $2.7\times10^{5}$ stars. Both photometries cover the main body of the galaxy, spanning roughly 2 square degrees on the sky in the $B$, and $V$ filters, and extending up to more than $\sim 1.2^\circ$ to the north-east from the centre of the galaxy. The field of view covered by the data is shown in Fig.~\ref{fig:Data}.\\

Since two different catalogues are being used, two aspects must be checked to ensure the homogeneity and self-consistency of the final photometry list: first, that all star magnitudes are in the same photometric system; and secondly that, since both catalogues overlap a large area in the sky, there are not repeated stars. Besides, the complete sample must be clean from non-stellar objects. In order to fulfil these requirements we followed the procedures described in Appendix~\ref{Ap:Internal_Calibration}.\\

\section{The CMD of Fornax}\label{Cap:CMD}

\begin{figure}
\begin{center}
\includegraphics[trim=0.cm 0.3cm 0.cm 0.24cm, clip=true, scale=0.75]{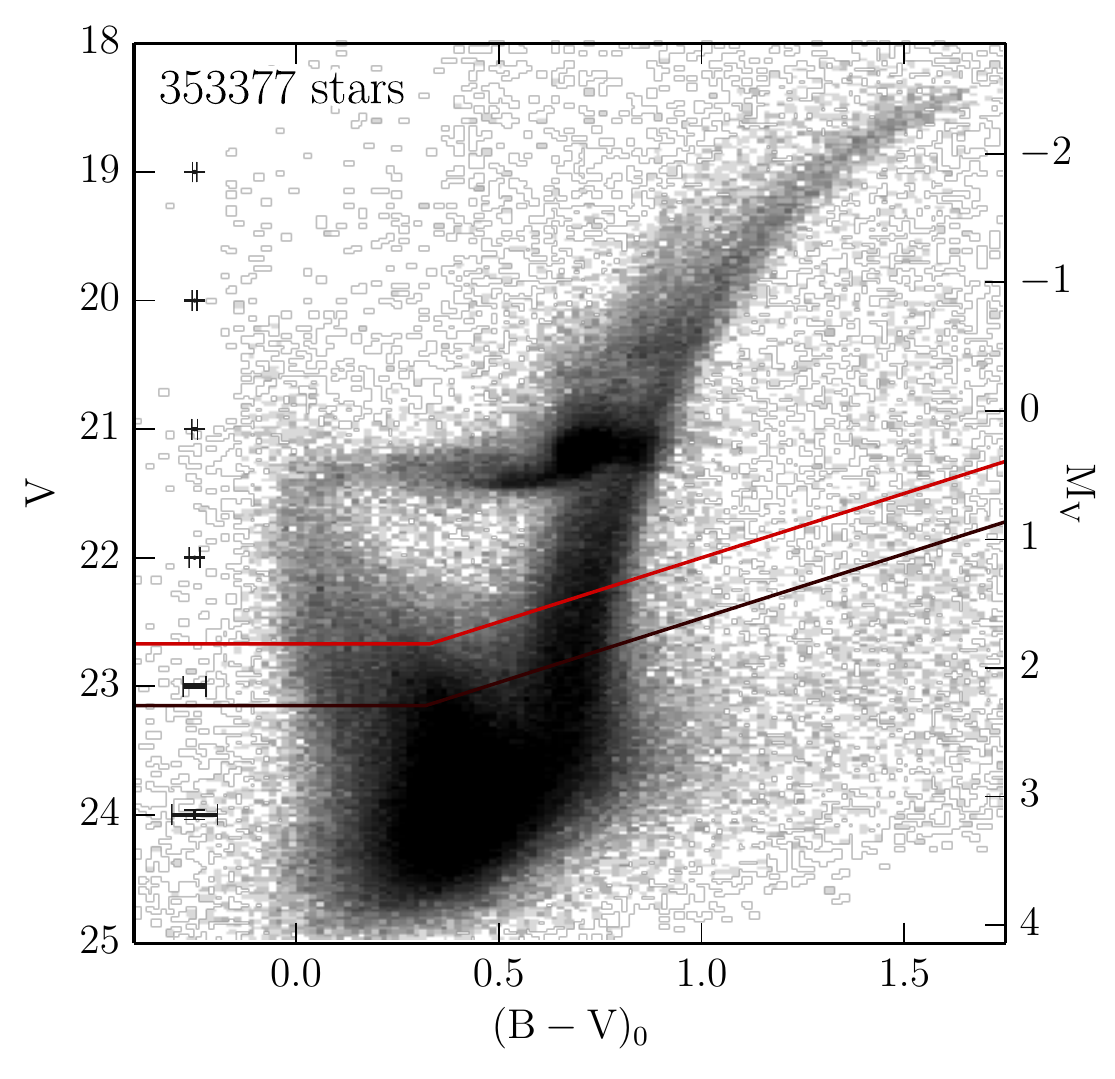}
\caption[Wide field CMD]{Hess diagram of the calibrated CMD of Fornax. Shading indicates the stellar density in logarithmic scale. Photometric errors appear as error bars in the left side of the figure. The completeness levels of 80 and 60 per cent for the innermost regions of the galaxy are shown by a light, and a dark red line respectively (see Fig.~\ref{fig:Completeness}).}
\label{fig:CMD}
\end{center}
\end{figure}

The calibrated CMD of Fornax is shown in Fig.~\ref{fig:CMD}. The most conspicuous features are a young main sequence (MS), and strongly populated red giant branch (RGB) and red clump (RC). The bluest and brightest stars of the MS ($M_V\leq 0$, $(B-V)_0\sim -0.1$) are younger than 1 Gyr. The RGB is formed by intermediate-age and old stars, older than 1--2 Gyr. Its thickness, significantly larger than the photometric error, suggests a spread in metallicity. The RC is centred at $(B-V)_0=$ 0.7 and $M_V=$ 0.5. It is essentially populated by helium-burning stars, with relatively large metallicities, and young-to-intermediate ages. Also a large number of very old, metal-poor stars ($> 10$ Gyr, $Z\simeq 0.001$) populate the horizontal branch (HB). The latter indicates the presence of low mass stars older than 10 Gyr. Finally, a slight over-density is apparent approximately 0.5 mag above the RC, identified as the asymptotic giant branch (AGB) clump. This wide spectrum of features indicates a complex SFH.\\

\subsection{Sampling the photometry}\label{Cap:CMD:CMDsample_1}

Stars arrange in a CMD according to their intrinsic characteristics (i.e. mass and metallicity) and to their evolutionary state. Stars born within the time interval $[t,t+{\rm d}t]$ and within the metallicity interval $[Z,Z+{\rm d}Z]$ can be formally defined as a simple stellar population. Therefore, a more complex stellar population can be considered as an ensemble of several simple stellar populations, which can be studied in detail by means of stellar evolutionary models. This makes out the CMD a powerful tool, providing valuable information about the different stellar populations of a stellar system through comparison with stellar evolutionary models.\\

\begin{figure}
\begin{center}
\includegraphics[trim=0.cm -0.22cm 0.cm 0.cm, clip=true, scale=0.75]{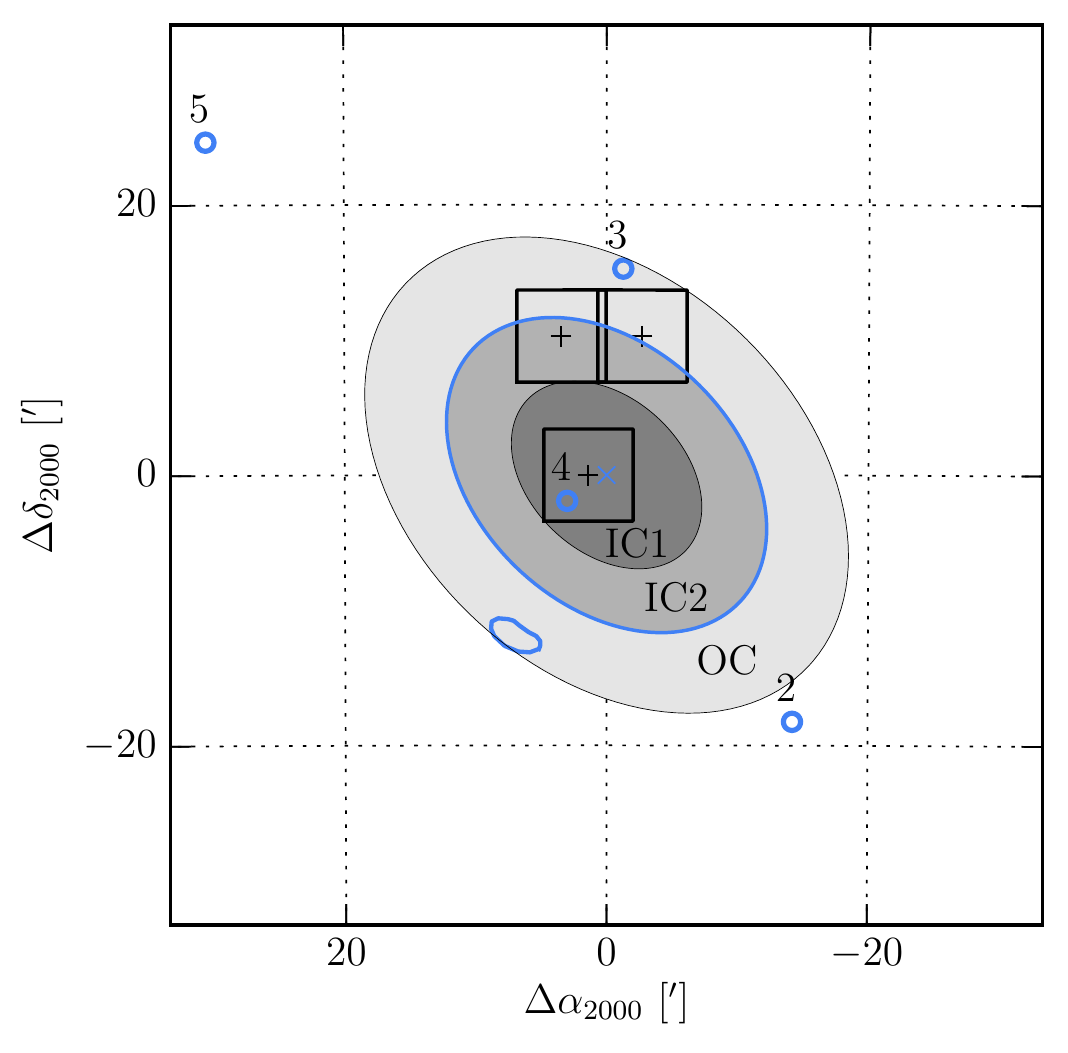}
\caption[Adopted axisymmetrical model for deriving the global SFH]{Axisymmetric radial model adopted for estimating the global SFH of Fornax. Fields in which the SFH was derived are represented as black squares, while their centres appear as plus symbols. Different coloured annular regions show the regions in which the SFHs of IC1, IC2, and OC were extrapolated. Markers coincide with those used in Fig.~\ref{fig:Data}.}
\label{fig:Averaged_Radial_model}
\end{center}
\end{figure}

\begin{figure}
\begin{center}
\includegraphics[scale=0.55]{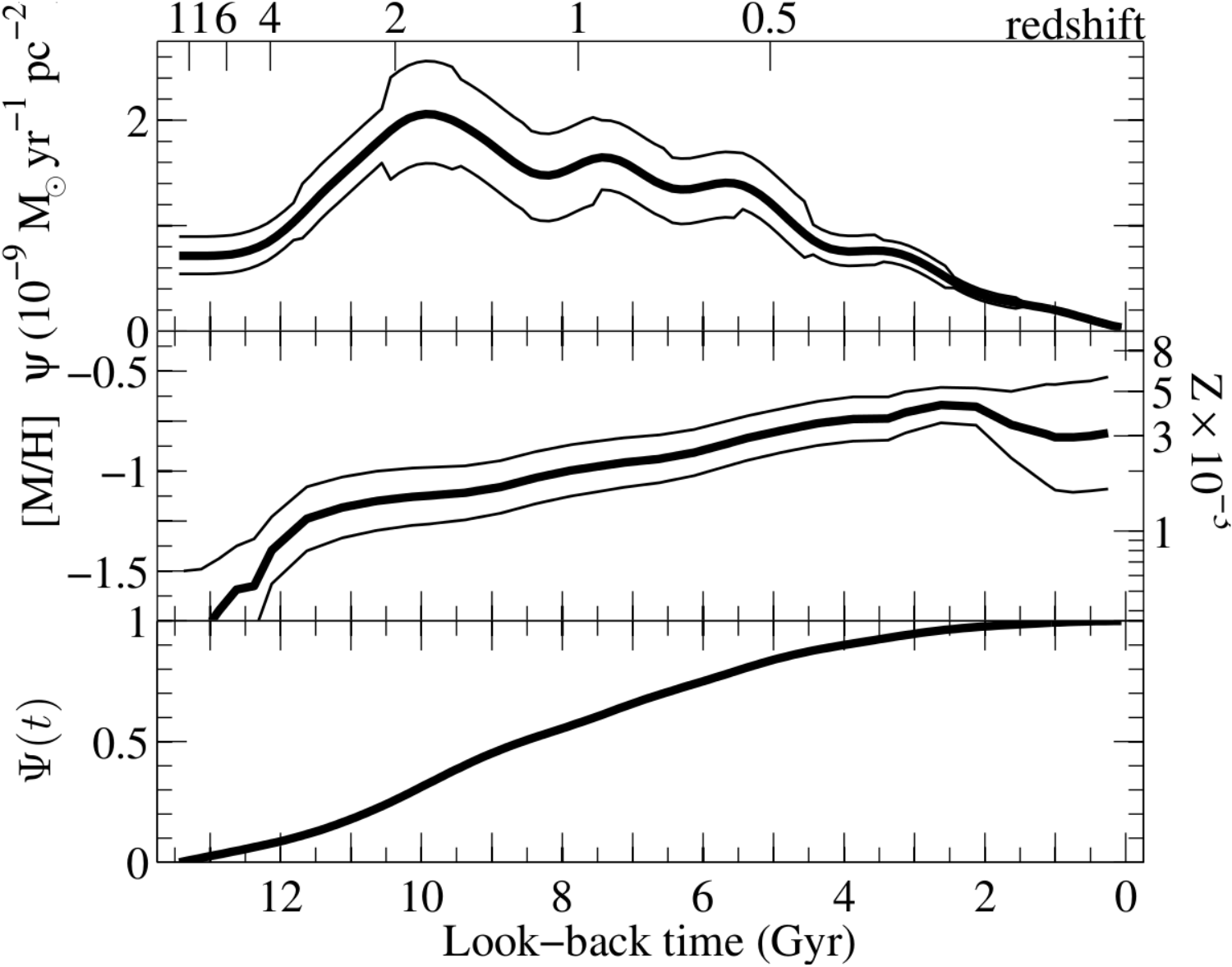}
\caption[The SFR, AMR, and cumulative mass fraction as a function of time for the global SFH]{The star formation rate as a function of time ($\psi(t)$) (top), the AMR (middle), and the cumulative mass fraction, $\Psi(t)$, (bottom) of the radial modelled SFH of Fornax. Error bands in $\psi(t)$ and dispersion in the AMR are drawn by thin lines. Units of $\psi(t)$ are normalized to the total area covered by the model. A redshift scale is given in the upper axis, computed assuming $H_0=70.5$km $\times$ s$^{-1}$Mpc$^{-1}$, $\Omega_m$=0.273, and a flat universe with $\Omega_\Lambda$=$1-\Omega_m$.}
\label{fig:Fornax_model}
\end{center}
\end{figure}

In order to extract the maximum information from the observed CMD of Fornax we used \textsc{IAC-Star} \citep{Aparicio2004} to compute a synthetic CMD (sCMD) based on the averaged SFH derived in \citet{delPino2013}, i.e., using the age-metallicity relation (AMR) and the SFR derived for Fornax as inputs for \textsc{IAC-Star}. The resulting sCMD corresponds to a stellar population representative of the average SFH of Fornax, computed as the arithmetic mean of the three SFHs derived within IC1, IC2, and OC regions, weighted by the areas of the elliptical annuli centred in Fornax which enclose each one of these regions \citep[see][]{delPino2013}. This sCMD can be sampled, providing information about the properties of the observed stars. In Fig.~\ref{fig:Averaged_Radial_model} we show the radial model adopted, while in Fig.~\ref{fig:Fornax_model}, we show the corresponding averaged SFH of Fornax used for synthesizing the sCMD. Some relevant integrated and averaged quantities associated with it are summarized in
Table~\ref{tab:Integrated_Quantities_2}.\\

Other ingredients of the \textsc{IAC-star} code used for generating the sCMD were the following: the scaled-solar BaSTI \citep{Pietrinferni2004} stellar evolution library, completed by \citet{Cassisi2007} models for very low mass stars, a fraction of 40 per cent of binaries, and the initial mass function by \citet{Kroupa2002}, i.e., a power law with exponent $x=2.3$ for stellar masses above $\rm 0.5~M_\odot$, and $x=1.3$ for stellar masses between 0.08--0.5 $\rm M_\odot$.\\

\begin{table}
  \caption{Integrated Quantities derived from the radial modelled SFH of the Fornax dSph.}
  \label{tab:Integrated_Quantities_2}
 \centering
  \begin{tabular}{@{}lccc}
    \hline
    \hline
      $\int{\psi(t'){\rm d}t'}$ \footnote{Integrated between 0 and 13.5 Gyr.} & $\langle\psi(t)\rangle$  & $\langle\rm age\rangle$ & $\langle Z \rangle$ \\
      $[10^7$ M$_\odot]$ & $[10^{-9}$ M$_\odot \rm yr^{-1} pc^{-2}]$ & $\rm[Gyr]$ & $\times10^{-3}$\\
      \hline
      $2.19 \pm 0.06$   &  $1.08 \pm 0.03$  & $8.23 \pm 0.8$ & $2.0 \pm 0.4$\\
    \hline
  \end{tabular}
\end{table} 

\textit{Observational effects} play a crucial role when deriving intrinsic properties of resolved stellar populations. These include stellar crowding, signal-to-noise limitations, detector defects and all factors affecting and distorting the observational data with the resulting loss of stars, changes in measured colours and magnitudes, and systematic uncertainties. These effects must be simulated in the sCMD before performing any quantitative analysis of it.\\

The most precise way to quantify the observational effects is through artificial star tests \citep[see][for an example]{Hidalgo2011}. However, this process requires large computer resources besides the considerable effort of re-obtaining the photometry from the original images, while the information which it would provide is beyond present work scope. For this reason, we made the observational effects simulation using a different, simpler approach.\\

For each filter, a boxcar function is applied over the photometric errors of the observed stars covering the full range of magnitudes, providing discrete functions of the average photometric errors and their dispersion as a function of the observed magnitudes. Results are fitted by a third-order spline, obtaining continuous functions from which the expected average photometric error can be obtained as a function of the magnitude. The photometric error for a specific synthetic star is then calculated, in each filter, through a simple random sampling of a normal distribution centred in the corresponding photometric error for its magnitude and using the corresponding photometric error dispersion as its standard deviation. Finally, this shift is randomly added or subtracted from the original magnitude for each synthetic star and filter. Despite being this error simulation not as precise as the one achieved through artificial star tests, it is sufficient to our current objectives, allowing realistic enough comparisons between the observed and synthetic CMDs.\\

The aforementioned comparison of CMDs also requires a smart sampling of the diagrams, since it is important to obtain detailed information from the well-known stellar evolutionary phases while avoiding the regions of the CMD for which the stellar evolution theory provides less accurate information or those that are affected by large observational effects.\\

\begin{figure*}
\begin{center}
\includegraphics[scale=0.75]{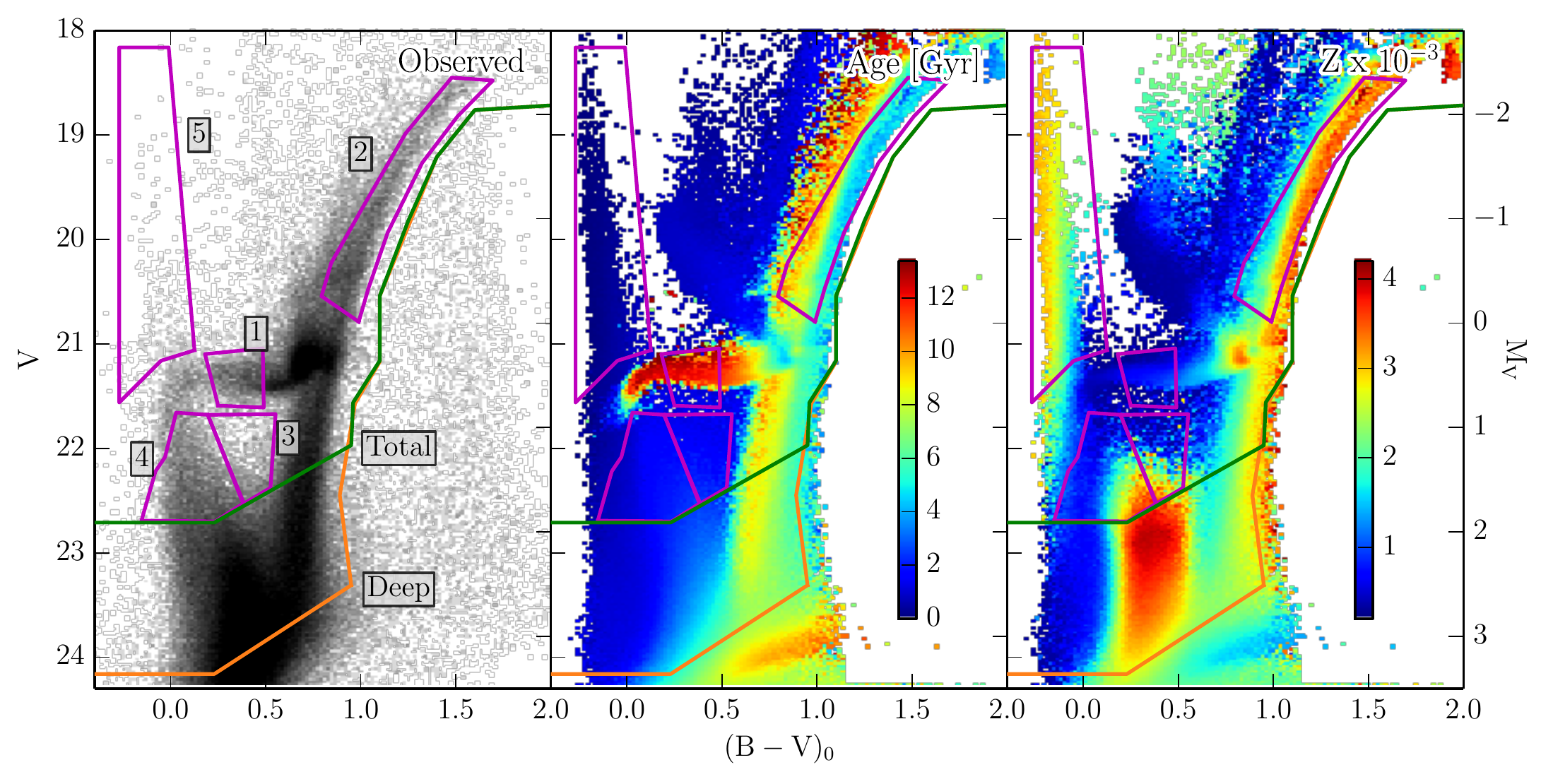}
\caption[CMD, sCMD and \textit{bundles} for the wide field photometry]{CMD of Fornax and the defined regions for the study of the spatial distribution of the stars (\textit{bundles}). Left-hand panel shows the Hess diagrams of the calibrated CMD of Fornax (see Fig.~\ref{fig:CMD}), while middle and right ones show the Hess diagrams of the sCMD with the expected average age, and the expected average metallicity, respectively. \textit{Bundles} are shown in all panels, and labelled accordingly in the left-hand panel.}
\label{fig:Stars_Selection}
\end{center}
\end{figure*}

The sCMD with simulated errors was used as a template to define five macro-regions in the observed CMD, delimiting the evolutionary phases used for our analysis. The regions, \textit{bundles}, are shown in Fig.~\ref{fig:Stars_Selection}. The sampling is as follows: (1) HB, (2) RGB, (3) the subgiant branch (SG) stars, (4) the young MS stars, and (5) the youngest stars of the MS (YMS). In all cases, bundles were created above the $\sim$80 per cent completeness level in the observed CMD (see Fig.~\ref{fig:Completeness}).\\

The same bundles were later applied to the observed CMD, defining five subpopulations in Fornax, each one covering a relatively narrow age range (except for the RGB which comprises stars from $\sim2$ to 13.5 Gyr). The averaged quantities measured over the stars within each bundle are listed in Table~\ref{tab:Bundles}.\\

\begin{itemize}
\item Column 1: bundle number;
\item Column 2: CMD region (in the following we will use the names listed here to refer to each subpopulation);
\item Column 3: mean age and $\rm \sigma(age)$ of sCMD stars;
\item Column 4: mean metallicity (Z) and $\rm \sigma(Z)$ of sCMD stars;
\item Column 5: number of observed stars, N, with their counting Poissonian error,$\rm \sqrt{N}$;
\item Column 6: non-Fornax sources contaminants. See Appendix~\ref{Ap:Internal_Calibration:Cleanning_Background}.
\end{itemize}

\begin{table}
  \caption{Basic data of the six defined macro regions for the CMD sampling.}
  \label{tab:Bundles}
  \begin{tabular}{@{}lccccc}
    \hline
    \hline
    \# & Subpop. & $\rm \langle Age \rangle$ & $\rm \langle Z \rangle$ & $\rm N_\star$ & Contaminants\\
           &               & $\rm [Gyr]$               &  $\times{10^-3}$          &               & $\rm [N_{sources}/(^\prime)^2]$\\
    \hline
    1 & HB   & $12\pm3$    & $0.7\pm0.2$ & 3060 $\pm$ 50   &  $0.02\pm0.01$ \\
    2 & RGB  & $7\pm3$     & $2.4\pm0.9$ & 9400 $\pm$ 100  &  $0.02\pm0.01$ \\
    3 & SG   & $1.7\pm0.4$ & $2\pm1$ & 1650 $\pm$ 40   &  $0.04\pm0.02$ \\
    4 & MS   & $1.1\pm0.7$ & $1\pm1$ & 7480 $\pm$ 90  &  $0.07\pm0.02$ \\
    5 & YMS  & $0.2\pm0.2$ & $2.2\pm0.6$ & 450  $\pm$ 20  &  $0.006\pm0.006$ \\
    6 & Total& $6\pm4$     & $2\pm1$ & 84100 $\pm$ 300 &  $0.57\pm0.06$ \\
    \hline
  \end{tabular}
\end{table} 

\subsection{Re-sampling the photometry: Averaged quantities}\label{Cap:CMD:CMDsample_2}

Not all the evolutionary phases of a star provide information of the same accuracy, nor the stellar evolution library data are equally accurate along the evolutionary tracks. In order to improve the age and metallicity resolution, we re-sampled both CMDs including fainter stars. We defined a big bundle similar to the one used for the total population, but reaching $M_V = 3.45$, close to the oldest main sequence turn off (oMSTO). This bundle was then divided into approximately 6144 boxes distributed in 96 $(B-V)$ colour and 64 $V$ magnitude bins. Given a real star within a box, its luminosity, mass, age, and metallicity are calculated as the averages of these quantities for the synthetic stars within the same box, while the standard deviation of such quantities is assumed to be their errors.\\

This information allows us to obtain completeness-corrected, high-resolution distribution maps of the luminosity, mass, age, and metallicity of the stars currently present in the galaxy with magnitudes as deep as $M_V = 3.45$.\\

\section{Two dimensional distribution maps}\label{Cap:2DMaps}

\subsection{Spatial distribution of subpopulations}\label{Cap:2DMaps:Desity_distribution}

\begin{figure*}
\begin{center}
\includegraphics[scale=0.75]{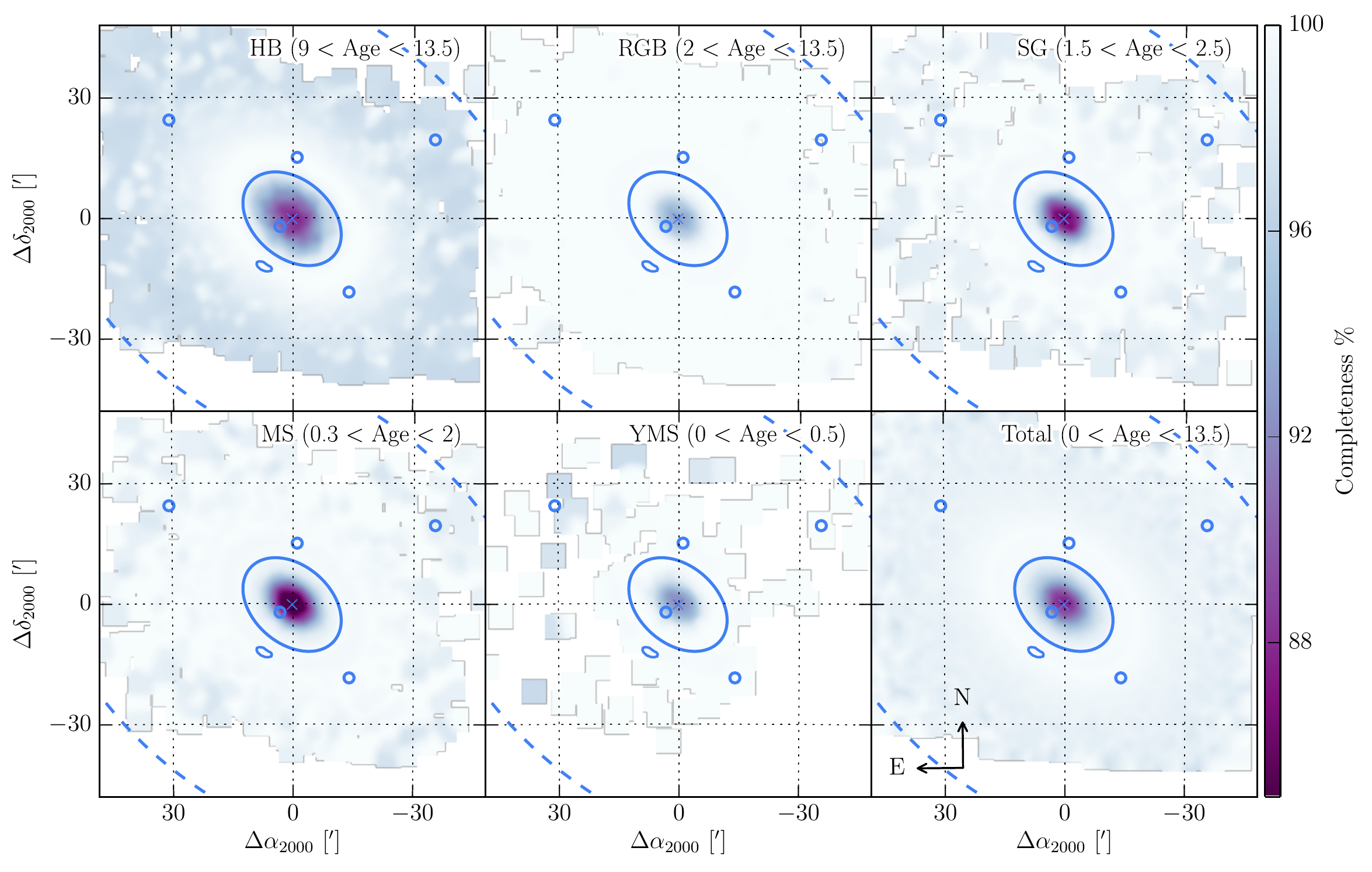}
\caption[Completeness maps]{Average completeness maps for the five defined subpopulations in the CMD. Panels are ordered from old to young populations, and labelled with the approximate age range. Colour scale is common for all panels, and indicates the completeness level. Markers coincide with those used in Fig.~\ref{fig:Data}.}
\label{fig:Completeness_Maps}
\end{center}
\end{figure*}

\begin{figure*}
\begin{center}
\includegraphics[scale=0.75]{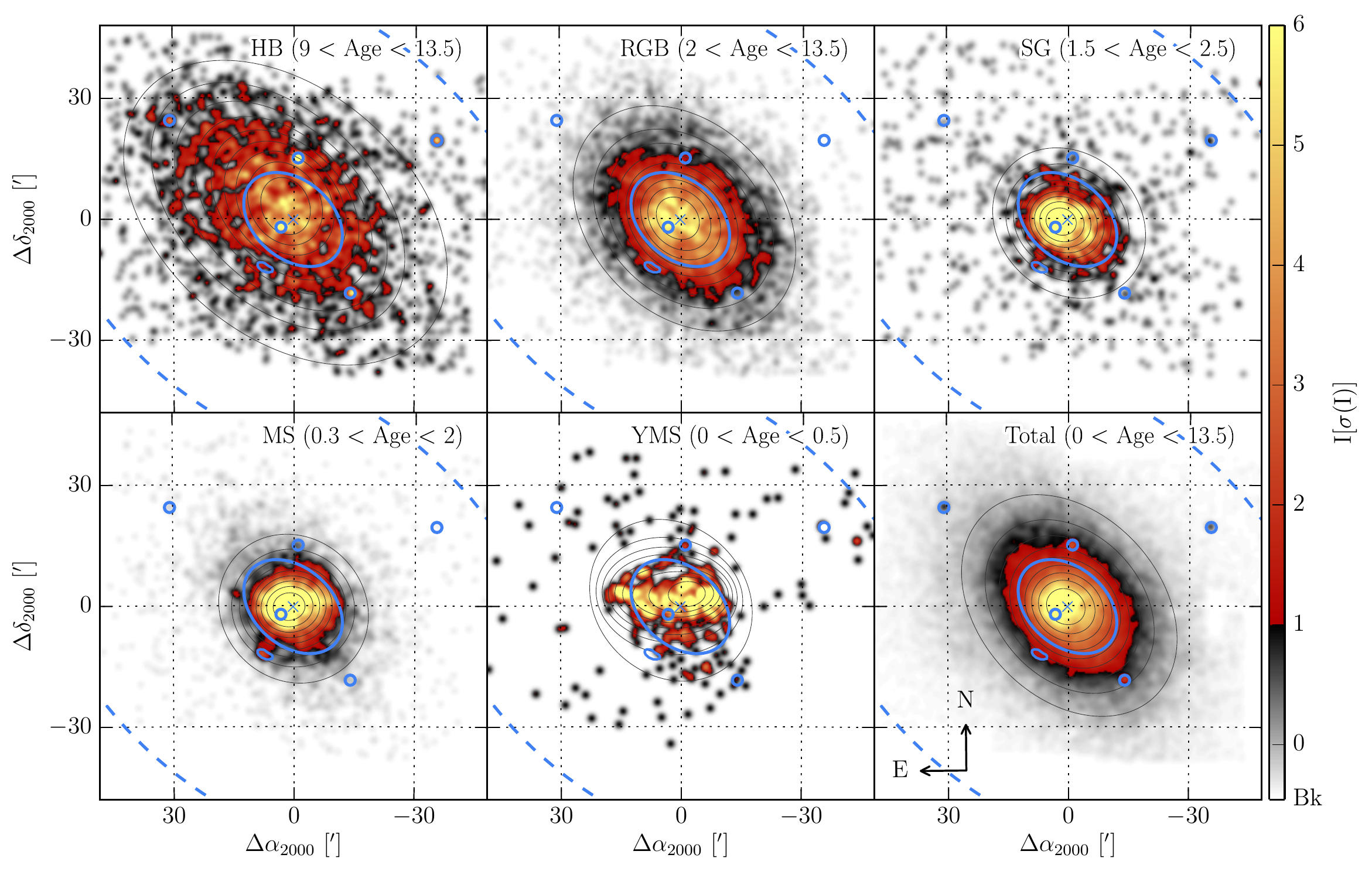}
\caption[Surface density maps]{Spatial distribution maps for the five defined subpopulations in the CMD. Panels are ordered from old to young populations, and labelled with the approximate age range. Colour scale is common for all panels, and indicates the concentration of stars normalized to the number of stars of each subpopulation. Levels are in units of standard deviation for each subpopulation. Black narrow-lined ellipses represent the obtained elliptical boundaries for deriving the radial density profiles (see Section~\ref{Cap:Radial_Profiles:Ellipse_fitting}). Markers coincide with those used in Fig.~\ref{fig:Data}.}
\label{fig:Density_Maps}
\end{center}
\end{figure*}

We have created surface density maps for each subpopulation defined by the bundles (see Fig.~\ref{fig:Stars_Selection}). These were obtained using $512\times512$ pixel two-dimensional histograms over the spatial distribution of the stars within each bundle. These histograms were convolved with a Gaussian kernel of 3 pixel width in order to enhance the most important features avoiding stochastic noise. Completeness maps were created following the same procedure and using the completeness associated with each star (see Appendices~\ref{Ap:Internal_Calibration:Data_Completeness} and \ref{Ap:Internal_Calibration:Data_Completeness:Spatial_Correction}). The completeness maps are shown in Fig.~\ref{fig:Completeness_Maps}. The surface density maps were divided by the completeness maps in order to obtain surface density maps corrected from completeness. These are shown in Fig.~\ref{fig:Density_Maps}.\\

\begin{figure*}
\begin{center}
\includegraphics[scale=0.75]{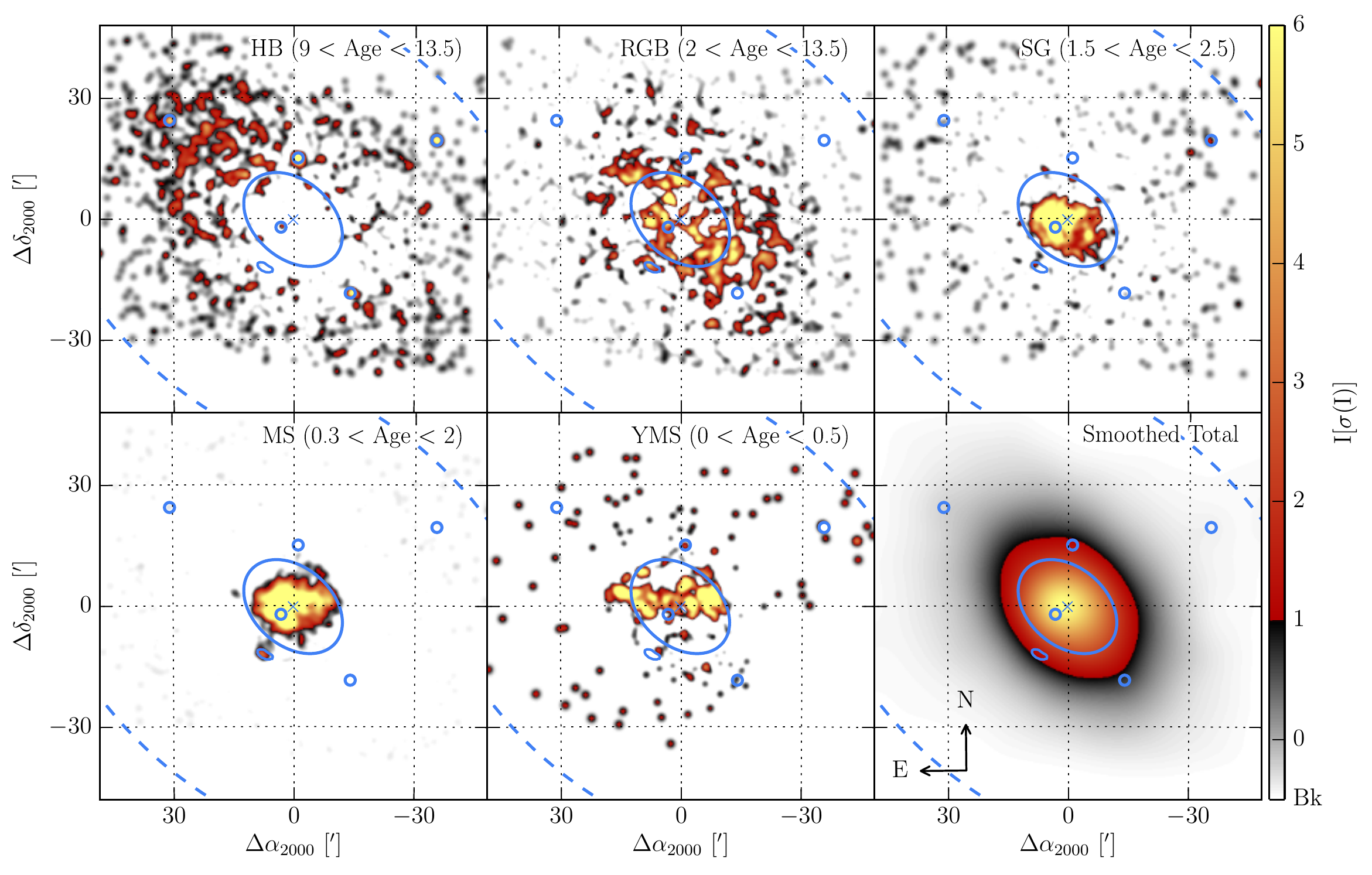}
\caption[Unsharp-masked surface density maps]{Unsharp-masked version of Fig.~\ref{fig:Density_Maps}.}
\label{fig:Density_Maps_Subtracted}
\end{center}
\end{figure*}

Differences between subpopulations are evident. Stars older than $\sim$ 1.5 Gyr follow an elliptical distribution, usually observed in dSph galaxies, while younger stars show distributions which do not follow the major axis of the core ellipse. They concentrate in the centre-most regions of the galaxy, populating also the inner shell structure found by \citet{Coleman2004}. The youngest stars of Fornax ($< 0.6$ Gyr), form an elongated or stream structure from east to west in the centre of the galaxy. This morphology could explain the presence of the two overdensities located at the ends of the feature at $\Delta\alpha_{2000} \sim 15^\prime, \Delta\delta_{2000} \sim 5^\prime$ and $\Delta\alpha_{2000} \sim -10^\prime, \Delta\delta_{2000} \sim 2^\prime$, which could be the projected positions of the stars in an edge-on elongated structure. The east-most of these clumps was reported by \citet{deBoer2013} as an isolated overdensity.\\

In order to enhance the features present in each subpopulation, we have computed the unsharp-masked surface density maps from the density maps presented in Fig.~\ref{fig:Density_Maps}. These were obtained by subtracting a very smoothed (Gaussian kernel of 15 pixel) surface density map of the total population to each subpopulation map, all of them previously normalized to their average density. This procedure enhances features present in subpopulations outstanding above the average distribution of stars in Fornax. The unsharp-masked surface density maps are shown in Fig.~\ref{fig:Density_Maps_Subtracted}.\\

It is clear that young stars gather in the central regions of the galaxy. This can be seen, for example, in the oldest star distribution map ($>$ 9 Gyr), with densities lower than the average in the centre, but not in the outskirts, where old stars clearly dominate the normalized density spectrum. Also noticeable is a slight overdensity of these old stars at the north-east semi-major axis of the galaxy.\\

Interestingly, in these maps the shell-like clump is also visible in the RGB. This will indicate that the age of at least some of the stars populating the clump are $\gtrsim$2 Gyr.\\

These differences in the spatial distribution of the stellar populations show a segregated structure and suggest a tumultuous past for Fornax. Moreover, some asymmetries can be observed depending on the age of the stars. Such asymmetries are not expected in a system in dynamical equilibrium in which all stars have been formed from the same molecular cloud in a monolithic collapse. Strongest asymmetries appear to be present in populations younger than 5 Gyr. This does not exclude the possibility of existing asymmetries in older populations: the large age-metallicity degeneracy of the RGB stars may be blending any specific distribution feature making it difficult to observe. The elliptical distribution of the RGB stars may arise from the contributions of different spatial distributions of stellar populations from $\sim$2 to $\sim$13.5 Gyr. Considering stars hosted in a triaxial potential well, the superposition of all of them should follow an elliptical surface distribution. Asymmetries in Fornax are widely known, but mostly discussed over filtered photographic plates or shallow photometry. \citet{Hodge1961} found an apparent 
asymmetry along the major axis of the galaxy which was later corroborated by
\citet{Eskridge1988a, Eskridge1988b, Irwin1995}, and \citet*{Stetson1998}.\\

Yet more intriguing is the fact that the shell-like structure reported by \citet{Coleman2004}, and \citet{Coleman2005} appears to be part of the main distribution of MS stars (see the MS distribution map in Fig.~\ref{fig:Density_Maps}). This suggests that the overdensity could be in fact a sky-projected stream of stars aligned with the line of sight. Stars of this stream would be still falling towards the Fornax potential well centre, moving in a plane defined from south-east to north-west and in the line-of-sight direction. However, this conclusion should be taken with caution since precise proper motions and line-of-sight velocities of these stars are required in order to confirm this scenario.\\

When comparing the centres of the different subpopulations spatial distributions, some of them show important offsets from the optical centre. SG, and MS stellar distributions are displaced towards the globular cluster Fornax 4. This effect was also found by \citet{Demers1994}. Only the RGB star distribution appears to be well centred on the optical centre of the galaxy derived by \citet{van_den_Bergh1999}. The RGB bundle samples stars with masses above $\sim 0.8 \rm M_\odot$ and with ages above $\sim 2$ Gyr, being these the brightest stars in Fornax ($\rm V \leq 20.5$). This match between the optical and RGB distribution centres is therefore logical, since the latter is the brightest subpopulation. Nevertheless, taking into account the SFH derived in \citet{delPino2013}, and adopting the commonly used mass-luminosity exponent $a=3.5$ \citep{Harwit1988}, low-mass MS stars may dominate the mass spectra of Fornax. The lack of depth of our photometry makes impossible to study \textit{in situ} the spatial distribution of these old low-mass stars. Nevertheless, we can assume that they follow the same distribution shown by the HB stars, of the same age. This should be taken into account when considering the centre of mass of the whole system, since it is expected to be shifted from the optical centre towards the barycentre of the much more numerous, older low-mass stars. In Table~\ref{tab:Ellipses_promedio} we list the most important parameters of the fitted ellipses, measured at the core radius.\\

\subsection{Averaged quantities distribution maps}\label{Cap:2DMaps:Averaged_distribution}

Using the information, derived in Section~\ref{Cap:CMD:CMDsample_2}, we have computed the luminosity, mass, age, and metallicity maps of Fornax. Following a similar procedure to the one described in the previous Section~\ref{Cap:2DMaps:Desity_distribution}, the averaged magnitudes maps have been obtained by averaging the relevant quantities of the stars in $512\times521$ pixels 2-dimensional grids. The dispersion of each magnitude, together with the completeness of the stars, was used as weight for the averaging. For each pixel $(i,j)$, its corresponding magnitude was calculated as:
$$
\langle {\rm M} \rangle_{ij} = \frac{\sum_{k=0}^{N}{{\rm M}_k\sigma({\rm M}_k)^{-1}C_k^{-1}}}{\sum_{k=0}^{N}{\sigma({\rm M}_k)^{-1}C_k^{-1}}}
$$
where ${\rm M}_k$ is the quantity (age, $Z$, mass, or luminosity) of the $k^{th}$ star, $\sigma({\rm M}_k)$ is its dispersion, $C_k$ is the completeness of the $k^{th}$ star, and $N$ is the number of stars within the $(i,j)$ pixel. In the case of the integrated luminosity, and total mass per pixel, their values were calculated as
$$
{\rm M}_{ij} = \sum_{k=0}^{N}{{\rm M}_k\sigma({\rm M}_k)^{-1}C_k^{-1}}
$$

\begin{figure*}
\begin{center}
\includegraphics[scale=0.75]{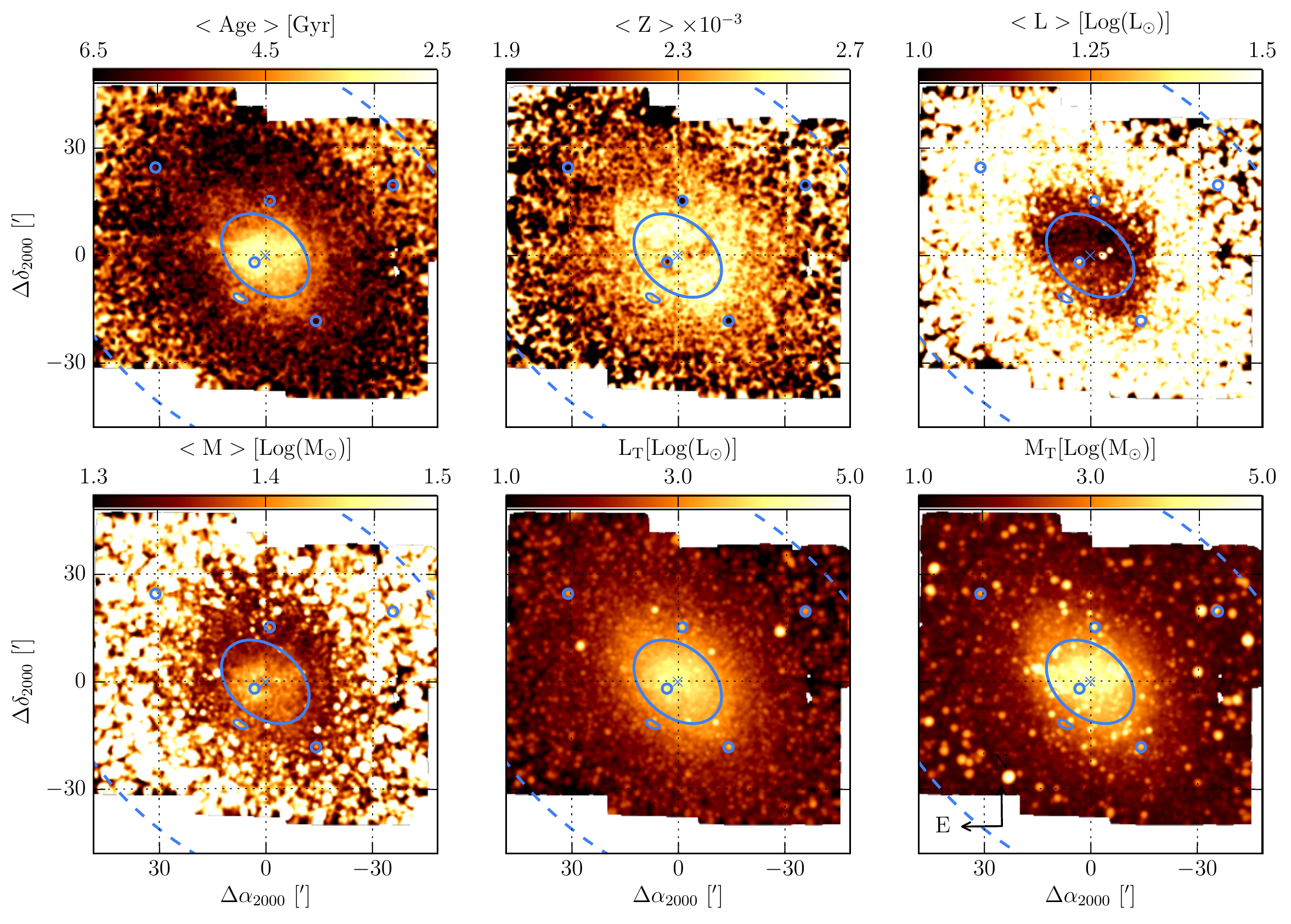}
\caption[Averaged distributions maps]{Mean distribution map of the sampled CMD. Colour scale represents the averaged magnitudes measured over the stars in the deep bundle (see Fig.~\ref{fig:Stars_Selection}). From left to right and top to bottom, the average age [Gyr], average metallicity $Z\times10^{-3}$, average luminosity [log(L$_\odot$)], and average mass [log(M$_\odot$)] of the stars are shown. The two bottom-right panels show the integrated luminosity and integrated mass distributions in logarithmic colour scale. Markers coincide with those used in Fig.~\ref{fig:Data}.}
\label{fig:Age_Map}
\end{center}
\end{figure*}

Maps with the distribution of average age, metallicity, luminosity, mass, as well as the integrated mass and luminosity are shown in Fig.~\ref{fig:Age_Map}. These show the completeness corrected averaged quantities of the observed stars, down to $M_V = 3.45$ as a function of the position.\\

Results are consistent with those of Fig.~\ref{fig:Density_Maps}. In the age panel, a decrease in the average age of the stars can be clearly observed within the core of the galaxy. These young stars form conspicuous features like the shell-like structure, and the clump located at $\Delta\alpha_{2000} \sim 15^\prime$, $\Delta\delta_{2000} \sim 5^\prime$, which clearly stand out as almost isolated features. Besides these, more irregular structures can be noticed. The most important ones are a small arc of young stars located at the south-west, outside the core of the galaxy ($\Delta\alpha_{2000} \sim -10^\prime$, $\Delta\delta_{2000} \sim -13^\prime$), and some substructure at the north-west part of the galaxy ($\Delta\alpha_{2000} \sim -5^\prime$, $\Delta\delta_{2000} \sim 10^\prime$).\\

A significant number of young stars gather near the globular cluster Fornax 4 ($\Delta\alpha_{2000} \sim 3.6^\prime$, $\Delta\delta_{2000} \sim -1.9^\prime$). These form an arc-like structure extending in the north--south direction (see the mass panel). The distribution in metallicity is not well aligned with the body of the galaxy, appearing to be, in average, more metal-rich the stars in the north-west regions. Besides, clumps of low-metallicity stars can be observed within the core of the galaxy, near Fornax 4 and at $\Delta\alpha_{2000} \sim -7^\prime$, $\Delta\delta_{2000} \sim 5^\prime$.\\

Interestingly, there are not strong correlations between age and metallicity maps, indicating a large metallicity dispersion in the centre-most regions of the galaxy. This is consistent with the one observed in the AMR of Fornax for stars younger than 3 Gyr \citep{Piatti2014}.\\

Asymmetries can be also observed by looking at the average-mass and average-luminosity-per-pixel panels. The aforementioned arc-like structure shows a relative higher luminosity, and mass, which indicates that it is formed by young massive stars. The shell like structure stands out in the mean age panel, and in the mean mass panel, thus indicating the position of the youngest and most massive stars in Fornax.\\

\section{The radial density profiles}\label{Cap:Radial_Profiles}

Several previous works have analysed the radial distribution of stars in Fornax \citep[e.g.]{Eskridge1988a, Irwin1995, Battaglia2008a}. Models can be fitted to these profiles providing important constrains about the size and mass of the galaxy.\\

\subsection{Ellipse fitting}\label{Cap:Radial_Profiles:Ellipse_fitting}

To obtain radial density profiles of each subpopulation we fitted ellipses to their completeness corrected surface density maps. Several criteria can be used to fit these ellipses. In order to make the minimum assumptions about the shape or the extension of the distribution of the stars we proceeded as follows. First we applied a broad Gaussian filter (25 pixels) to the surface density maps to reduce noise fluctuations. Isopleths were computed for these maps from the maximum density level to the minimum one in steps of 10 per cent of the density range of each subpopulation (i.e. maximum to minimum density difference). Lastly, ellipses were fitted to these isopleths following the procedure of \citet{Fitzgibbon1996}. Free parameters were the centre, both semi-major axis lengths, and the position angle (PA, $\theta$). A set of ten ellipses was created for each subpopulation, defining a set of nine elliptical regions which were later used for star counting. The final ellipse sets are shown in Fig.~\ref{fig:Density_Maps}. Ellipse parameters measured at the core radius per subpopulation are listed in Table~\ref{tab:Ellipses_promedio}. Columns provide the following information:\\

\begin{itemize}
\item Column 1: subpopulation;
\item Column 2, 3: coordinates of the centre of the ellipse fitted at the core radius ($\alpha_{2000}$, $\delta_{2000}$);
\item Column 4, 5: distance of coordinates given in columns 2, and 3 from the centre of Fornax defined by \citet{van_den_Bergh1999};
\item Column 6: core radius ($r_{\rm c}$). Estimated as the distance along the semi-major axis ($a$) at which the surface brightness has dropped by half;
\item Column 7: Average ellipticity $\epsilon = 1 - b/a$, where $b$ is the
  semi-minor axis and $a$ is the semi-major axis;
\item Column 8: Average position angle (PA), measured over the angles sustained by the semi-major axes with the north direction.
\end{itemize}

\begin{table*}
  \caption{Parameters of the fitted ellipses set for each subpopulation measured at the core radius.}
  \label{tab:Ellipses_promedio}
  \begin{tabular}{@{}lccccccc}
    \hline
    \hline
    Subpop.& $\alpha_{2000}$      & $\delta_{2000}$       & $\Delta\alpha_{2000}$ & $\Delta\delta_{2000}$ & $ r_{\rm c} $ & $\langle \epsilon \rangle$ & $\langle \theta \rangle$\\
           & $[^\circ]$           & $[^\circ]$            & $[^\prime]$ &$[^\prime]$ & $[^\prime]$ &  & $[^\circ]$\\    
    \hline
    HB     & $39.97\pm0.04$       & $-34.45\pm0.06$       & $ 0.62$ & $0.83 $ & $22\pm7$       & $0.32\pm0.1$    & $43\pm2$ \\
    RGB    & $40.0201\pm0.0004$   & $-34.510\pm0.004$     & $-0.44$ & $0.02 $ & $16\pm2$       & $0.23\pm0.04$   & $43\pm3$ \\
    SG     & $40.00\pm0.02$       & $-34.520\pm0.006$     & $-0.11$ & $-0.86$ & $10\pm4$       & $0.14\pm0.05$   & $64\pm2$ \\
    MS     & $40.01\pm0.01$       & $-34.509\pm0.004$     & $-0.16$ & $-0.32$ & $10\pm4$       & $0.10\pm0.09$   & $95\pm1$ \\
    YMS    & $39.97\pm0.04$       & $-34.45\pm0.05$       & $-1.35$ & $0.75 $ & $14.1\pm0.2$   & $0.4\pm0.2$     & $88\pm2$ \\
    Total  & $40.02958\pm0.00007$ & $-34.51417\pm0.00002$ & $0.46 $ & $-0.76$ & $13.92\pm0.07$ & $0.191\pm0.001$ & $46.15\pm0.06$ \\    
    \hline
  \end{tabular}
\end{table*}

\subsection{Ellipses morphology}\label{Cap:Radial_Profiles:Ellipse_morphology}

\begin{figure}
\begin{center}
\includegraphics[scale=.75]{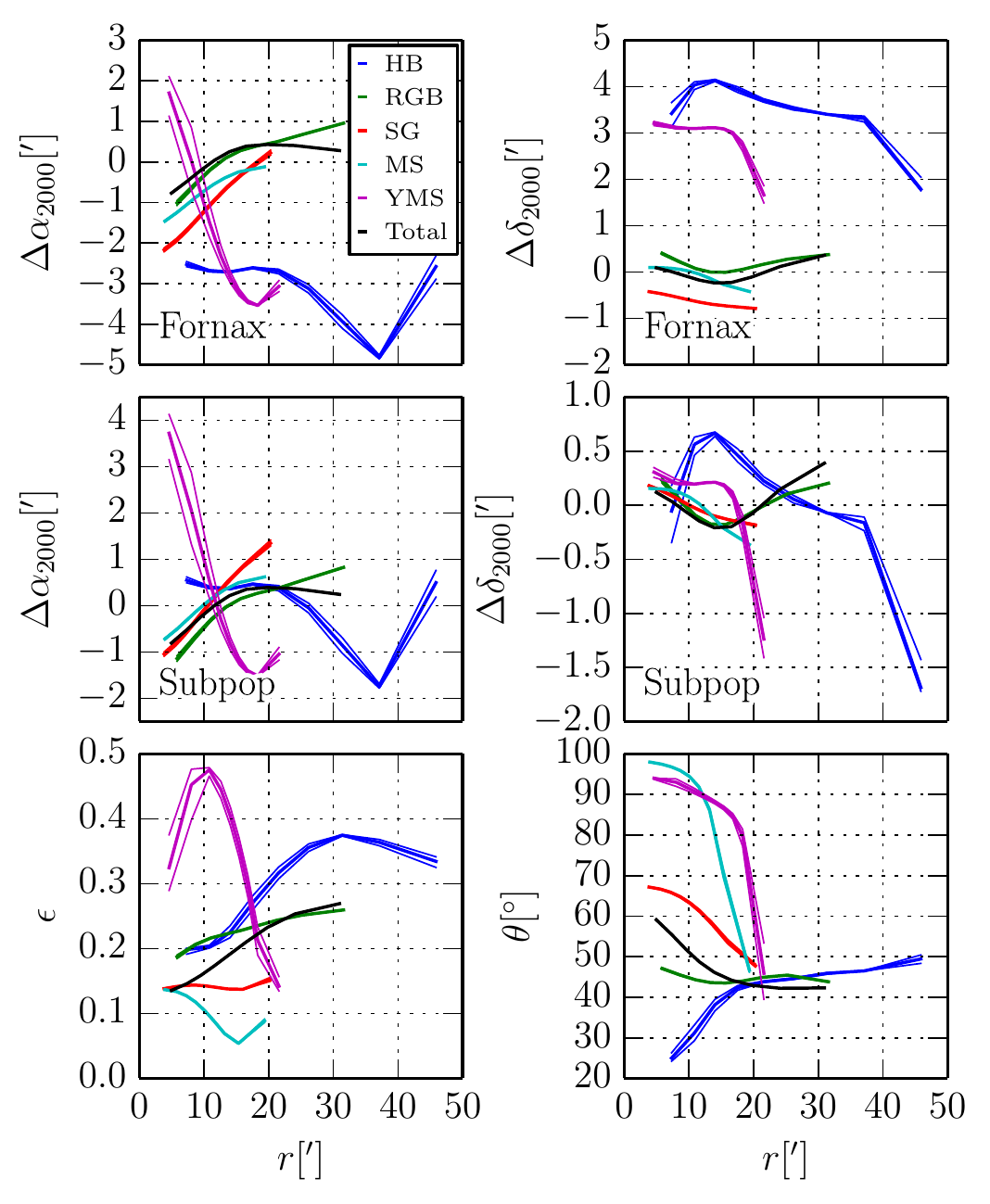}
\caption[Fitted ellipses parameters as a function of radius]{Evolution in the ellipse parameters as a function of the galactocentric radius. Top panels show the variation in right ascension and declination respectively with respect to the adopted centre of Fornax. Middle panels show same variation, but with respect to the barycentre of each subpopulation ellipse set. Ellipticity ($\epsilon$) and PA are shown in the bottom panels. Errors are shown by thin lines.}
\label{fig:Ellipses_evolution}
\end{center}
\end{figure}

Ellipses show an evolution in morphology as a function of both galactocentric distance and subpopulation. These trace broadly the spatial distribution of the stellar populations at different galactocentric distances. Fig.~\ref{fig:Ellipses_evolution} shows the changes of the ellipses parameters as a function of the semi-major axis length.\\

Interestingly, populations do not share the same baricentre. For small galactocentric distances, almost all populations appear to be centred slightly to the east, except the YMS stars, whose centre is located at the north-west of the galaxy. As we move to outer regions, all the intermediate-aged populations tend to stabilize their positions close to the optical centre of Fornax. This is not the case for the oldest (HB), and the younger (YMS), stars which are centred at the north-east for larger galactocentric distances. The ellipticity, $\epsilon$, also provides relevant information. For the older stellar populations (HB, RGB),  $\epsilon$ increases with the distance. Younger populations (SG, and MS) show a different behaviour, being their $\epsilon$ smaller. The youngest YMS population shows an increasing $\epsilon$ for small radial distances and a reversal of this tendency at approximately $10^\prime$ from their distribution centre. This could indicate some contamination from HB stars in the YMS distribution (see Fig.~\ref{fig:Stars_Selection}), since we do not expect many real young stars beyond $r = 10^\prime$. Possible effects of this contamination are expected to be negligible for $r < 10^\prime$, where young stars are much more numerous than blue, metal-poor HB stars.\\

A clear segregation between the stellar populations can be done by observing their PA for relatively small galactocentric radii. Older stellar populations show small PA, while it increases as we consider younger populations. Basically, older stars follow a rather stable spatial distribution, oriented roughly $41^\circ$ to the north-east. Younger stars, in contrast, follow an east-west distribution, with a high ellipticity in the case of the YMS population. This behaviour may indicate the presence of dynamical decoupled components in the galaxy.\\

\subsection{Fitting stellar surface density profiles}\label{Cap:Radial_Profiles:Fitting_profiles}

Using the set of ellipses defined in Section~\ref{Cap:Radial_Profiles:Ellipse_fitting}, we have obtained the radial density profiles of the stars in Fornax. These ellipses delimit a set of concentric elliptical regions in which stars were counted. Besides, a galactocentric distance must be adopted for each elliptical region in order to obtain the stellar density as a function of the radius. Since the ellipses do not share the same centre, we adopted this galactocentric distance to be the average distance of the stars to the subpopulation barycentre. This makes the equivalent radius of each elliptical region not directly related to the subpopulation centre distance. This method also assures a more realistic distance measurement of the stars to the centre, without supposing any particular structure in the galaxy. The stellar density as a function of radius for each subpopulation is listed in Appendix~\ref{Ap:Radial_Density_Profiles_Tables} tables.\\

Three models were fitted to the radial profiles: a  King model \citep{King1962, King1966}, a Plummer sphere model \citep{Plummer1911}, and a S\'ersic profile \citep{Sersic1968}. The King and Plummer are dynamical models for stellar self-gravitating systems. By identifying $I(r)$ with the number of stars per unit of area as a function of the radius, we can express the King model as:
$$I(r) = I_0\left[\frac{1}{[1+(r/r_{\rm c})^2]^{1/2}} - \frac{1}{[1+(r_{\rm t}/r_{\rm c})^2]^{1/2}}\right]^2$$
where $I_0$ is the stellar central surface density, $r_{\rm c}$ the core radius, and $r_{\rm t}$ the tidal radius.\\

The much simpler Plummer sphere model accounts for the total stellar mass of the system, $M$, and a characteristic scale length $\alpha$. By supposing the same mass for all the stars, $M$ becomes the total number of stars $N_\star$ of the system:
$$I(r) = \frac{\alpha^2N_\star}{\pi(r^2 + \alpha^2)^2}$$\\

The last fitted model, the S\'ersic profile, is a purely empirical model of the form
$$I(r) = I_0 \exp\left[-\left(\frac{r}{r_{\rm c}}\right)^{\frac{1}{n}}\right]$$
in which, as with the King model, $I_0$ and $r_{\rm c}$ stand for the central intensity and the core radius. The so-called S\'ersic index is $n$.\\

For each model, the fitting was performed in two different ways. The first one is an error weighted least squares fitting. The error of each parameter was obtained by multiplying the covariance matrix of the fit by the residual variance of the parameter. The second method accounts for the possible counting errors in the density profiles through extensive Monte Carlo experiments. A shift is introduced randomly in each profile point according to its Poissonian error (see Tables~\ref{tab:Radial_Profiles_annul1_Subpop} and \ref{tab:Radial_Profiles_annul1_Total}). The fitting is performed in the same way as before, but over this new data set with random shifts. The complete experiment is repeated $1\times10^{6}$ for each model and population, obtaining the corresponding sets of fitting parameters and their $\chi^2$. For each specific parameter, its final value and error are the median and the standard deviation of the whole distribution of possible values weighted with their $\chi^2$.\\

In some cases, the least-squares method resulted in a singular Jacobian matrix, which does not allow us to obtain the associated error. For this reason, we adopted the Monte Carlo results as the final parameters for each model.\\

\subsection{Derived parameters for models}\label{Cap:Radial_Profiles:Model_Parameter}

\begin{figure*}
\begin{center}
\includegraphics[scale=0.75]{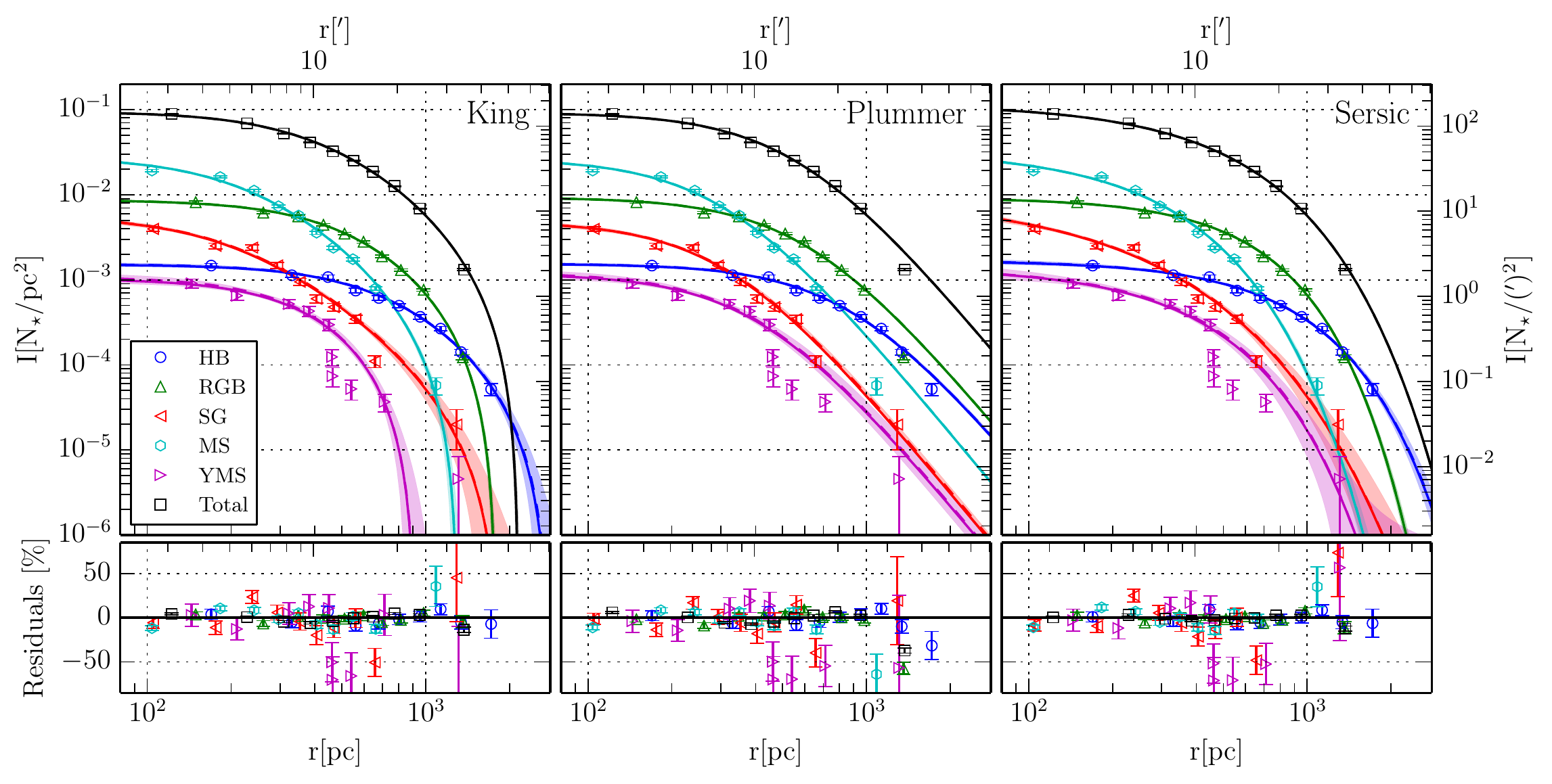}
\caption[Radial surface density profiles using free ellipse fitting]{Radial surface density profiles for each subpopulation. Measured within the elliptical regions defined in Section~\ref{Cap:Radial_Profiles:Ellipse_fitting} (see Fig.~\ref{fig:Density_Maps}). The three fitted models, King, S\'ersic, and Plummer are shown in panels from left to right, and are labelled accordingly. The best fits to these models are shown with continuous lines for the least squares approach, while those obtained through Monte Carlo inference are shown with dashed lines. Shaded areas show the $\pm1\sigma$ of the goodness of the fitting with monte carlo. Bottom panels show the residuals from the fit using least squares. Errors bars include the counting Poissonian error from the sources and from the contaminants subtraction.}
\label{fig:Radial_Profiles}
\end{center}
\end{figure*}

The radial profiles for each subpopulation and their fitted models are shown in Figure \ref{fig:Radial_Profiles}. The fitted parameters for the King, Plummer, and S\'ersic models are listed in Tables \ref{tab:King_model}, \ref{tab:Plummer_model}, and \ref{tab:Sersic_model} respectively. The format of the three tables is similar, except for the listed parameters:\\

\begin{itemize}
\item Column 1: subpopulation;
\item Column 2, from second to last: best fitting parameters through the monte carlo method.
\item Last column: resulting $\chi^2$ from the fitting.
\end{itemize}

\begin{table*}
  \caption{Fitting parameters for the King model for each subpopulation.}
  \label{tab:King_model}
  \begin{tabular}{@{}lccccccc}
    \hline
    \hline
    Subpop. & \multicolumn{7}{c}{King parameters} \\
            & \multicolumn{2}{c}{$I_0$} & \multicolumn{2}{c}{$r_{\rm c}$}  & \multicolumn{2}{c}{$r_{\rm t}$}                                 & $\chi^2$ \\
            & $\rm[pc^{-2}]\times10^{-2}$       & $\rm[(^\prime)^{-2}]$ & [pc]        &  [$^\prime$] & [pc]         &  [$^\prime$]  &        \\
    \hline
    HB      & $0.28\pm0.02$         & $4.4\pm0.3$                & $760\pm60$  & $19\pm1$     & $2800\pm200$ & $70\pm6$      & 1.05 \\
    RGB     & $1.81\pm0.04$         & $28.4\pm0.6$               & $580\pm20$  & $14.7\pm0.4$ & $1790\pm30$  & $45.2\pm0.7$  & 1.39 \\
    SG      & $0.70\pm0.05$         & $11.0\pm0.7$               & $200\pm20$  & $5.0\pm0.4$  & $1800\pm300$ & $47\pm8$      & 6.70 \\
    MS      & $4.02\pm0.09$         & $63\pm1$                   & $221\pm6$   & $5.6\pm0.1$  & $1310\pm40$  & $33.1\pm0.9$  & 7.11 \\
    YMS     & $0.3\pm0.3$           & $4\pm5$                    & $400\pm300$ & $10\pm8$      & $920\pm80$   & $23\pm2$     & 38.45 \\
    Total   & $14.5\pm0.1$          & $228\pm2$                  & $424\pm5$   & $10.7\pm0.1$ & $2160\pm20$  & $54.6\pm0.6$  & 13.51 \\
    \hline
  \end{tabular}
\end{table*} 

\begin{table*}
  \caption{Fitting parameters for the Plummer model for each subpopulation.}
  \label{tab:Plummer_model}
  \begin{tabular}{@{}lcccccc}
    \hline
    \hline
    Subpop & \multicolumn{4}{c}{Plummer parameters}\\
    & $N_\star$ & \multicolumn{2}{c}{$\alpha$} & $\chi^2$ \\
    &     $\times10^{4}$        & [pc] &  [$^\prime$]\\
    \hline
    HB    & $0.41\pm0.01$   & $920\pm20$ & $23.2\pm0.5$   & 2.10  \\
    RGB   & $1.16\pm0.01$   & $629\pm7$  & $15.9\pm0.2$   & 44.51  \\
    SG    & $0.163\pm0.006$ & $330\pm10$ & $8.2\pm0.3$    & 3.62  \\
    MS    & $0.83\pm0.01$   & $315\pm4$  & $8.0\pm0.1$    & 12.13  \\
    YMS   & $0.069\pm0.007$ & $430\pm40$ & $11\pm1$       & 23.24  \\
    Total & $9.67\pm0.04$   & $578\pm2$  & $14.61\pm0.06$ & 60.40  \\
    \hline
  \end{tabular}
\end{table*}

\begin{table*}
  \caption{Fitting parameters for the S\'ersic model for each subpopulation.}
  \label{tab:Sersic_model}
  \begin{tabular}{@{}lcccccc}
    \hline
    \hline
    Subpop       & \multicolumn{6}{c}{S\'ersic parameters}\\
                 & \multicolumn{2}{c}{$I_0$} & \multicolumn{2}{c}{$r_{\rm c}$} & $n$ & $\chi^2$\\
                 & $\rm[pc^{-2}]\times10^{-2}$ & $\rm[(^\prime)^{-2}]$ & [pc] &  [$^\prime$] &  &  \\

    \hline
    HB    & $0.17\pm0.01$ & $2.7\pm0.2$  & $700\pm40$ & $18\pm1$     & $0.73\pm0.05$   & 0.94 \\
    RGB   & $0.92\pm0.03$ & $14.4\pm0.5$ & $530\pm10$ & $13.3\pm0.3$ & $0.66\pm0.02$   & 1.91 \\
    SG    & $1.0\pm0.3$   & $15\pm4$     & $140\pm40$ & $3\pm1$      & $1.2\pm0.2$     & 6.22 \\
    MS    & $3.3\pm0.2$   & $52\pm3$     & $210\pm10$ & $5.4\pm0.3$  & $0.86\pm0.03$   & 6.71 \\
    YMS   & $0.14\pm0.04$ & $2.2\pm0.6$  & $310\pm60$ & $8\pm2$      & $0.8\pm0.2$     & 24.83 \\
    Total & $11.7\pm0.2$  & $183\pm3$    & $381\pm5$  & $9.6\pm0.1$  & $0.871\pm0.009$ & 6.35 \\

    \hline
  \end{tabular}
\end{table*} 

Results are consistent with each other between methods and models. S\'ersic profile proved to be the best fitting model overall, followed by King and lastly, the Plummer sphere, which failed fitting points at large radii. However, Plummer sphere model succeeded recovering the total number of stars sampled in each bundle. Looking at each subpopulation, it seems that oldest populations, like HB and RGB stars, better follow a King profile. Younger populations cannot be properly fitted by one-dimensional radial profiles, showing $\chi^2$ values over 5, and even 30 for the YMS stars. In any case, these were better fitted by S\'ersic profiles with $0.8 < n < 1.2$, indicating certain discy shape.\\

Basically, these important differences between the radial profiles account for the differences already stated in Section~\ref{Cap:2DMaps:Desity_distribution}. With $r_{\rm c} = 760\pm60$ pc, and  $r_{\rm t} = 2,800\pm700$ pc, the HB population is clearly more extended than the rest, showing a concentration parameter $[c = \log(r_{\rm t}/r_{\rm c})]$ relatively small, $0.56\pm0.05$, compared to the Total population, $0.72\pm0.03$. On the other hand, younger stars appear to be more concentrated in the central regions of the galaxy. This was also reflected by the S\'ersic profile and Plummer models. The S\'ersic index provides a clue about the shape of the population. All values for $n$, except for the SG are between 1 and 0.5, showing a relative dispersion of 22\%. This is the expected shape for a dSph and it indicates a rather flat central concentration of stars and truncation effect on the profile (an exponential profile).\\

The younger star distribution appears to be stochastic beyond some radius (see residuals in Fig.~\ref{fig:Radial_Profiles}). Bundle 5 was defined well above the magnitude corresponding to a completeness of 100\%, making unlikely these fluctuations being a consequence of bad sampling or completeness issues. Besides this, at the same star count levels, older subpopulations follow well the fitted models. The instabilities in the distribution of young stars are highlighted by comparing the YMS or SG density profiles with the HB one. This is even clearer if we take into account that bundle 1 is deeper than bundles 3 and 5 (see \ref{fig:Stars_Selection}). In summary, it is likely that younger stars profile model departures arise from real substructures present in their spatial distributions which do not follow an elliptical distribution.\\

In Table~\ref{tab:Fornax_final}, the structural parameters derived for the total population are listed. These were obtained using the information from the ellipse set measured at the core radius of the galaxy, together with the fitted parameters for the King model.\\

\begin{table}
\centering
  \caption{Main structural parameters derived for Fornax dSph.}
  \label{tab:Fornax_final}
  \begin{tabular}{@{}lc}
    \hline
    \hline
    Quantity & Value \\
    \hline
    RA, $\alpha$ (J2000.0) &  $\rm 2^h$ $\rm 40^m$ $\rm 7.1^s$ \\
    Dec., $\delta$ (J2000.0) & $-34^{\circ}$ $30^\prime$ $51^{\prime\prime}$ \\
    Ellipticity, $e$ & $0.20\pm 0.01$ \\
    PA $[^\circ]$ & $46.15\pm0.06$ \\
    Core radius $[^\prime]$ & $10.7\pm0.1$ \\
    Tidal radius $[^\prime]$ & $54.6\pm0.6$ \\
    \hline
  \end{tabular}
\end{table} 

\subsection{Comparison with previous works}\label{Cap:Radial_Profiles:Comparison_Previous_Works}

Results partially agree with those derived using other methods \citep{Eskridge1988a, Irwin1995, Battaglia2006, Walker2011}. We believe that differences from previous photometric studies can be accounted for by the fact that their photometries are shallower than ours. Therefore, they were tracing the position of the brightest RGB, YMS, and HB stars without the contribution of MS stars. In fact, values obtained in these works are much more similar to those obtained for the HB and RGB subpopulations rather than for the total population. Considering the existent differences between subpopulations, it is not clear, in our opinion, that a reliable measurement of the characteristic size of Fornax can be achieved by using relatively shallow photometry. It would be preferable to reach at least one magnitude below the oMSTO with good S/N, thus sampling the spatial distribution of all the stellar populations directly in the MS. Otherwise, a mass-follows-light spatial distribution model should be adopted, which is not to be the case for galaxies with asymmetries in their populations as Fornax.\\

\section{Interaction with other systems}\label{Cap:Interaction_Discussion}

Fornax is probably the most complex dSph galaxy satellite of the MW \citep{Stetson1998,Coleman2008,deBoer2012,delPino2013}. In the present paper, we have shown that its complex SFH reflects asymmetric distribution patterns in its stellar populations.\\

How Fornax has become such a complex dSph is still subject of debate. While some authors claim that this is the result of secular evolution processes \citep[e.g.][]{deBoer2013}, others think that it is related with past interactions with other systems \citep[see, for example,][]{Coleman2004,Amorisco2012,Yozin2012}. These interactions could have triggered star formation events, removed the gas, result in the accretion of gas, unbound stars from its potential well, etc. In this section, we analyse two possible interaction scenarios for Fornax, and their footprints in its actual properties.

\subsection{The MW influence}\label{Cap:Interaction_Discussion:MW}

Notwithstanding the relatively large perigalacticon distance (118$\pm^{19}_{52}$ kpc) and small orbit eccentricity \citep[0.3;][]{Piatek2007}, Fornax oldest populations exhibit several features that could have arisen as a consequence of tidal interaction with the MW: the ellipticity of the isopleths of the oldest stars increases with increasing galactocentric distance, reaching a maximum at around 32$^\prime$; the PA of ellipses changes with the galactocentric distance; isopleths show a larger gradient on the eastern side of the galaxy, and the map of the stellar surface density (Fig.~\ref{fig:Density_Maps}) shows stars of the dSph beyond the fitted tidal radius along the minor axis. It is worth investigating whether these features may be related to tidal stirring effects.\\

From fitted King profiles, it appears that the oldest stars of Fornax are well described by a relaxed, truncated isothermal model. Moreover, Fornax shows a bar amplitude of the Fourier decomposition of stellar phases $A_2 = 0.24\pm0.02$ for HB stars beyond the core radius, which can be related to mild tidal extensions. On the other hand, stars younger than $\sim$2 Gyr do not follow elliptical distributions, and cannot be fitted with 1D profiles (see Fig.~\ref{fig:Age_Map}).\\

According to \citet{Lokas2012}, it follows that more than two close passages are required in order to transform a disc-shaped system of the orbital parameters of Fornax into a spheroidal one. Assuming that perigalacticon passages occurred  at $\sim$2.6, $\sim$6.2, and $\sim$9.8 Gyr ago \citep[see][]{delPino2013}, it follows that the HB stars have suffered at least three close passages with the MW. An average SG star, in contrast, would have suffered only one at most, and MS, and YMS stars probably none.\\

This would suggest that, in fact, Fornax has suffered some tidal forces from the MW, which would have time enough to shape the oldest populations of the galaxy into a spheroidal system. Younger stars, on the other hand, did not suffer enough perigalacticon passages as to be transformed into a spheroidal component. Notwithstanding, we must emphasize that the calculation of these passages times has important uncertainties which increase with look-back time. These uncertainties are due to the large intrinsic errors in the orbital parameters derived by \citet{Piatek2007} as well as to the assumption of a very unrealistic model for the trajectory of Fornax during the last 12 Gyr.\\

\subsection{A merger scenario in Fornax}\label{Cap:Interaction_Discussion:Merger}

\subsubsection{Motivation}\label{Cap:Interaction_Discussion:Merger:Merger_Motivation}

The possibility of Fornax having suffered a merger has been already explored by many authors \citep[e.g.][]{Coleman2004, Yozin2012}. From the point of view of the spatial distribution of the stellar content, Fornax shows features which cannot be explained without strong dynamical events during its evolution. These events could include enhanced star formation bursts, SN feedback and mergers events.\\

Fornax shows strong spatial asymmetries when comparing its stellar populations, as well as shell-like structures and clumps of younger stars. It is also remarkable that, despite more than 10 Gyr of Fornax evolution, these young stars are not aligned with the main axes of the system. Being Fornax a kinematically supported system, the existence of main axes may indicate certain rotation signal, mainly around the minor axis of the system. This would be a good approximation for the younger populations, for which the MW influence is not expected to have had enough time as to transform their distributions. The misalignment of the young populations may indicate, therefore, a different rotation pattern with respect to the general trend. In fact, if young stars were rotating around Fornax, their momentum would be aligned to the north-south direction.\\

In this scenario, the possibility of material being accreted to Fornax naturally arises. In its fall into the Fornax potential well, this material would have acquired an angular momentum different from the predominant one in Fornax, forming structures not aligned with the galaxy main axes. Whether this subcomponent is stable and for how long depends on many factors, including whether the system can be approximated as collisionless, its secular evolution, etc. Indeed, Fornax can be safely considered as a collisionless system, with a relaxation time of $t_{\rm relax} \approx 10^2/{\rm H_0}$ (adopting a total mass of $10^8M_\odot$). From collisionless $N$-body experiments carried out for other galaxies \citep{Lokas2014}, it follows that anomalous rotation patterns can survive for several Gyr ($\lesssim$ 8 Gyr) if it is the dominant feature. Substructures like shells and streams from the merger can survive for a few dynamical times, flying around or through the system at least once but not many times, which for dwarf galaxies mean they would exist for less than 1 or 2 Gyr.\\

The fact that these substructures remain present in Fornax would indicate a very recent accretion of gas, less than 1--2 Gyr ago. Stars forming the shell-like structure are older than 1.5 Gyr, and younger than 2.5 Gyr (see Fig.~\ref{fig:Density_Maps}), which is consistent with the expected age of the substructure. Supposing that stars were born after this accretion, their age should be within 1.5 and 2 Gyr. This is in good agreement with results obtained by \citet{Olszewski2006}, who found an approximate age of 1.4 Gyr for the stars in the shell.\\

The origin of the gas from which these stars were formed remains unknown in any case. Three possible scenarios can be summarized as follows.\\

\begin{itemize}
 \item A cloud of fresher molecular gas, accreted by Fornax. Later, new stars were born from this gas conserving the angular momentum of the gas cloud.
 \item Infall of enriched gas previously expelled by Fornax due to stellar winds or SN feedback.
 \item A merger with other system, whose stars were accreted by Fornax.
\end{itemize}

All these are plausible scenarios which cannot be completely ruled out on the basis of our data. Nevertheless, assuming  the metallicity derived by \citet{Olszewski2006} for the shell stars ($[Fe/H] \sim -0.7$) makes the first scenario very unlikely since a pre-enrichment of the gas would be necessary to produce stars of such high metallicity. Besides, the lack of gas in Fornax makes improbable that these stars came from gas already present in the galaxy.\\

The metallicity of the shell stars appears to be in fact higher than the average metallicity of Fornax ($[Fe/H] \sim -0.95$), and it seems realistic to suppose that the gas from which these stars were born was pre-enriched in Fornax itself. Fornax could have suffered an outflow of gas related to strong stellar formation events or by the SN feedback. This gas would have been recaptured later producing the substructures we observe today. Being this true, the gas should have been expelled around $\sim 3$ Gyr ago, time from which the AMR of Fornax starts to show a metallicity around $[Fe/H] \simeq -0.7$. This scenario, already proposed by \citet{deBoer2013}, is supported by the small increment of star formation observed in the SFH $\sim$ 3 Gyr ago. Yet, we noted a tail of high metallicities ($[Fe/H] \gtrsim -0.4$) at $\sim8$ Gyr \citep{delPino2013}. Therefore, the main burst occurred in the central region of the galaxy 8 Gyr ago could have expelled an important amount of gas recaptured $\sim2$ Gyr ago. This scenario, could explain the substructures, and their apparent high metallicity, but not the origin of the high metallicity tail.\\

The results obtained here are also qualitatively compatible with a scenario in which rather small systems had been accreted. However, from cosmological $\Lambda$CDM simulations it follows that this would be very unlikely \citep{Mayer2010}. Due to their velocities, and the number of dwarf galaxies surrounding their host, mergers between satellite galaxies were very unlikely much later than $z\sim1$. This, together with the SFH results, encouraged us to propose a scenario in which Fornax suffered a major merger at redshift $z\sim1$ ($\sim$ 8 Gyr). This scenario was already introduced by \citet{delPino2013} and \citet{Piatti2014} inspired by the delay in the main burst in the central regions of the galaxy. Moreover, our analysis revealed a tail of stars with ages about 8.5 Gyr, and metallicities extending up to $[Fe/H] \gtrsim -0.4$. This metallicity spread clearly does not follow the general trend of the AMR of Fornax, with metallicities much higher than the average metallicity at $z\sim 1$.\\

\subsubsection{Merger footprints}\label{Cap:Interaction_Discussion:Merger:Merger_footprints}

The imprint of this merger should be visible in the stellar populations of Fornax. In fact, features suggesting such event have been already reported by several works \citep[i.e.][]{Amorisco2012, delPino2013, Hendricks2014}. Supposing that Fornax is the result from a major merger between two smaller systems, the high metallicity tail found in its AMR $\sim8.5$ Gyr ago could arise from a much more efficient chemical enrichment process in one of the progenitors. Being a rather small galaxy, the lack of gas would have fostered this rapid enrichment.\\

On the other hand, if the merger occurred at $z\sim 1$, Fornax would have had enough time to well mix the two gas components. We think that during the merger process an important amount of gas could have been expelled from the Fornax progenitors. Once the merger finished, the resulting galaxy, with a strengthened potential well, would have started recapturing the previously expelled gas. Since the gas retained in Fornax during the merger would have followed a chemical enrichment process, the inflow of the previously expelled gas (with lower metallicity) would be causing the larger metallicity spread observed in the last $\sim 2$ Gyr \citep[see][]{Piatti2014}.\\

The observed shell-like structures could have arisen from the inflow, $\sim$2 Gyr ago, of a large clump of this expelled gas. This would explain the relatively high metallicity of the stars populating the inner structure \citep{Olszewski2006}, since the infalling  gas was previously enriched until the merged occurred. The proposed scenario could also explain how Fornax has managed to keep forming stars until very recently, and the observed differences between the average-age and the average-metallicity maps (Fig.~\ref{fig:Age_Map}). These could have arisen from the presence of young stars born from not well-mixed gas.\\

The proposed scenario does not exclude the possibility of more mergers. From these results it follows that stars from the inner shell-like structure are between 1 and 2 Gyr old. Nevertheless, there are stars as young as 300 Myr (see Figures~\ref{fig:Density_Maps} and \ref{fig:Age_Map}). This longstanding star formation could be also compatible with the accretion of more gas or even stars from other systems very recently.\\

\subsubsection{What about the progenitors?}\label{Cap:Interaction_Discussion:Merger:Merger_progenitors}

\begin{figure}
\begin{center}
\includegraphics[scale=.65]{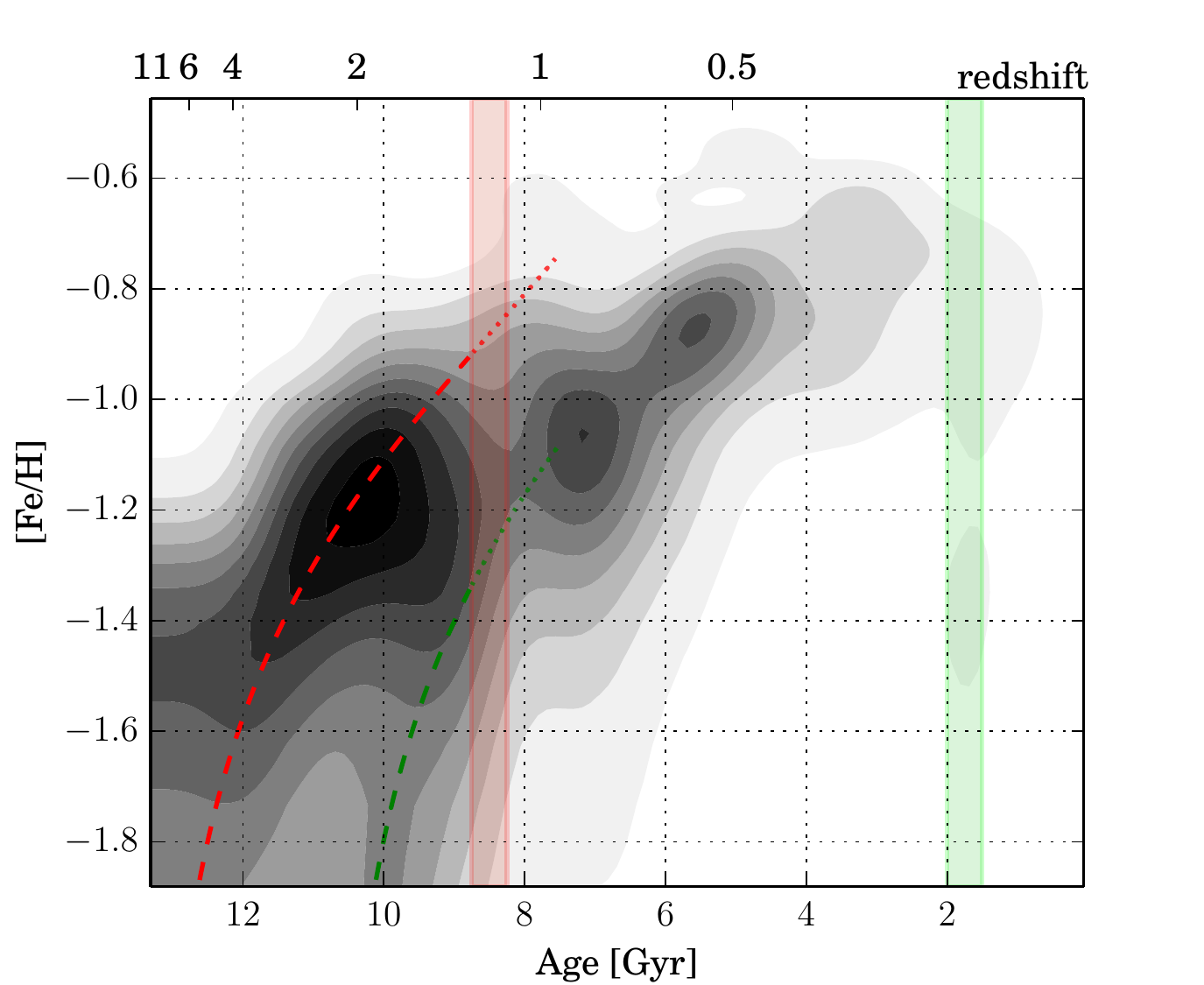}
\caption[Global SFH of Fornax and two outflow enrichment models]{Hess diagram of the average SFH for Fornax. Metallicity is given in logarithmic scale. Grey shades indicate the amount of mass converted into stars. The red and green regions indicate the moments in which the major merger and the accretion of the gas cloud which produced the stream would have occurred. The red and green dashed lines represent two toy AMRs constructed with outflow models. Both share the same parameters, except for the moment of starting to generate stars, 11 and 13.5 Gyr, respectively, and the relative masses.}
\label{fig:SFH_2D}
\end{center}
\end{figure}

In Fig.~\ref{fig:SFH_2D} we show the averaged SFH for the three regions studied in \citet{delPino2013}. Two toy enrichment models with constant SFR have been over-plotted for illustrative purposes. The green dashed line represents an outflow model in which the system started forming stars 10.75 Gyr ago, having consumed 55 per cent of its gas mass at 7.5 Gyr ago. The red one is also an outflow model starting 13.3 Gyr ago, with the 75 per cent of its gas consumed 7.5 Gyr ago. For simplicity, both models share the same \textit{effective yield} parameter of 0.0028. The red system has $\sim 93$ per cent of the mass of the green one. Supposing that these two systems were the progenitors of Fornax, and assuming that Fornax lacks gas at present, then approximately 60 per cent of the Fornax stars would have born before 7.5 Gyr ago. This is in very good agreement with the cumulative mass fraction derived for the galaxy, $\Psi(t)$ (see Fig.~\ref{fig:Fornax_model}), provided that Fornax had recovered most part of the gas expelled during the merger. One may note that the metallicity spread after the accretion at $\sim$ 2 Gyr is remarkably large, probably as a consequence of the presence of not well-mixed gas out of which the new generation of stars has been formed. Interestingly, the spread extends down to $[Fe/H] \sim -1$, which was roughly the mean metallicity of the galaxy $\sim$ 8 Gyr ago.\\

We cannot venture to make precise calculations about the total stellar mass of the progenitors and the amount of gas expelled, although we can provide some confidence intervals. From the SFH of the galaxy, we calculated the total mass ever converted into stars as $(2.19 \pm 0.06)\times10^7 M_\odot$. This is based on the radial model adopted up to a galactocentric distance of $20.7^\prime$ ($\sim 820$ pc). Supposing that both outflow models represent well the progenitors of Fornax, then their stellar masses would have been $M_{\star1} \approx (1.14\pm0.03)\times10^7 M_\odot$ and $M_{\star2} \approx (1.05\pm0.03)\times 10^7 M_\odot$. By supposing the inflow of the expelled gas from $\sim$ 7 Gyr ago to $\sim$2 Gyr, values in the range $0.02\times M_{\rm Gas}(t=0) < M_{\rm Gas\hspace{2pt} Expelled} < 0.35\times M_{\rm Gas}(t=0)$ appear to be reasonable. The latter provides a range between $5\times 10^5 M_\odot < M_{\rm Gas Expelled} < 8\times 10^6 M_\odot$.\\

Other possible scenario could be the accretion of a primeval gas clump roughly 2 Gyr ago. This inflow would have quenched the chemical enrichment process of Fornax. Being the average metallicity of Fornax $[Fe/H] = -1.0\pm 0.1$ at $z\sim 1$ (8 Gyr ago) and $[Fe/H] = -0.68\pm 0.09$ 2.1 Gyr ago, it follows that $\sim$ 48 per cent of the total gas converted into stars in Fornax during last 2 Gyr was accreted. Therefore, the total mass accreted would be$(2.44\pm0.07)\times 10^5 M_\odot$.\\

This is comparable to the stellar masses of Canes Venatici I, Draco, Sextans, and Ursa Minor, galaxies with total dynamical masses of the order of $10^8 M_\odot$. This mass is too small to make a late collision probable, even assuming that all the gas was accreted in the last 2 Gyr $[(5.1\pm0.1)\times 10^5 M_\odot]$. This suggests that the gas cloud was already bounded to the Fornax potential well, which strengthens the merger scenario.\\

These results are compatible with those of \cite{Amorisco2012} who, using line-of-sigh velocities, found different rotation patterns depending on the considered metallicity. They concluded that Fornax is the result of a late merger of a bound pair of galaxies. Moreover, they estimated the total luminosity of the accreted companion to be $L \approx 7 \times 10^5 L_\odot$, which coincides with the luminosity expected for a dSph galaxy in the range of masses we are handling.\\

\section{Summary and Conclusions}
\label{Cap:Summary}

Fornax has proved to be a very complex system, with a segregated structure. Here we list the main outstanding results of this work.

\begin{itemize}
 \item Oldest stars (10--13.5 Gyr) appear to follow a truncated spheroidal distribution, extending far beyond the core radius of the RGB stars.

 \item Young stars ($\lesssim 2$ Gyr) are concentrated within the central regions of the galaxy. They show strong asymmetries, besides conspicuous clumps of stars not aligned with the optical axes of the system.
  
 \item  Isopleth contours of the stellar surface density show important variations as a function of the galactocentric distance. This may indicate a possible tidal interaction of Fornax with the MW.
 
 \item King, Plummer, and S\'ersic profiles have been fitted to the data, obtaining their one-dimensional fitting parameters. Only older stars can be properly modelled by one-dimensional profiles. Oldest stars ($\gtrsim 9$ Gyr) have $r_{\rm c} = 760\pm60$ pc, while the youngest ($\lesssim 3$ Gyr) range from $140\pm40$ to $310\pm60$ pc. The RGB population has $r_{\rm c} = 580\pm20$ pc, being this result in good agreement with previous ones obtained by other authors.
 
 \item The shell-like structure reported by \citet{Coleman2004} appears to be in fact the projection in the sky of a stream of stars. This suggests the infall of material into Fornax with a different angular momentum than the one sustained by the oldest stars. 
\end{itemize}

We propose a major merger occurred at $z\sim 1$ as the most plausible scenario for explaining all these features. During this merger, an important amount of gas would have been expelled from the galaxy, being this recaptured in the subsequent Gyr by the resulting strengthened potential well. We have estimated the mass of the accreted cloud to be $(2.44\pm0.07)\times 10^5 M_\odot$. The recaptured cloud of gas $\sim$2 Gyr ago would be responsible for the streams of stars found in the younger populations and for the high-metallicity dispersion. This scenario does not exclude the possibility of more interactions of Fornax with other systems.\\

\section*{Acknowledgements}
 
The authors thank the referee, Dr R. Ibata, for his thorough review and comments. We are grateful to Dr P. Stetson and Dr. T. de Boer for fruitful discussions about the completeness of the data, and to Prof. E. {\L}okas for her suggestions regarding stellar density profile models. AdP also thanks S. Bertran de Lis for her inestimable support and contribution to this work. This research was partially supported by the IAC (grant 310394), and the Education and Science Ministry of Spain (grants AYA2010-16717, and AYA2013-42781-P). Further support has been provided by the Polish National Science Centre under grant 2013/10/A/ST9/00023.\\

%%%%%%%%%%%%%%%%%%%%%%%%%%%%%%%%%%%%%%%%%%%%%%%%%%



%\bibliographystyle{mnras}


%%%%%%%%%%%%%%%%%%%%%%%%%%%%%%%%%%%%%%%%%%%%%%%%%%

%%%%%%%%%%%%%%%%% APPENDICES %%%%%%%%%%%%%%%%%%%%%

\appendix

\section{Internal photometric calibration}\label{Ap:Internal_Calibration}

A very precise calibration between both catalogues, Cat1 and Cat2, is required in order to study the CMD of Fornax. We performed such calibration using Cat1 as the reference catalogue. This decision ensures consistent comparisons with results obtained in \citet{delPino2013}.\\

A first inspection of the common stars of both catalogues revealed small gradients in magnitudes between them. These should be corrected before undertaking further analysis. The internal calibration between catalogues was obtained comparing the magnitudes of the brightest and well measured stars of Cat1 with those of Cat2. The best measured stars were selected from each catalogue on the basis of their magnitudes, their photometric errors, the number of measurements per star, and quality flags provided by \textsc{Daophot} for Cat1 and \textsc{Dophot} for Cat2. More specifically, we only used stars with magnitudes within $\rm 18.5<V<23.5$ and a maximum photometric error of $\sigma=0.05$, in both filters and catalogues. With this selection, stars close to saturation as well as those with low signal-to-noise are avoided. Typically, several independent measurements were made for Cat1 stars. We only included stars whose average magnitude was obtained from at least five measurements in both filters $(B, V)$ in the reference catalogue. In the case of Cat2, only isolated stars were used for the calibration. This selection was performed on the basis of the quality flag $q$ provided by \textsc{Dophot}, which indicates if the star is isolated or it was recovered from the superposition of two or more stars. The calibration was performed through an iteratively least-squares sigma-clipped fitting as follows.\\

Both photometric catalogues were matched using \textsc{Stils} \citep{STILS}. Different criteria may be adopted in the matching process. As starting point, we matched stars using only their sky position. Stars closer than $0.65^{\prime\prime}$ were considered to be the same. This first matching provides a starting first order calibration between catalogues of the form:\\

\begin{math}
   B_{\rm Cat1} = a + b {\rm RA}_{\rm Cat2} + c {\rm Dec}_{\rm Cat2} + d B_{\rm Cat2} + e (B_{\rm Cat2}-V_{\rm Cat2})
\end{math}\\

\begin{math}
   V_{\rm Cat1} = a + b {\rm RA}_{\rm Cat2} + c {\rm Dec}_{\rm Cat2} + d V_{\rm Cat2} + e (B_{\rm Cat2}-V_{\rm Cat2})
\end{math}\\

\noindent where $(B_{\rm Cat1}, V_{\rm Cat1})$ are the calibrated magnitudes from Cat1, $(B_{\rm Cat2}, V_{\rm Cat2})$ the magnitudes from Cat2, and $({\rm RA}_{\rm Cat2}, {\rm Dec}_{\rm Cat2})$ the sky coordinates from Cat2. The fitting coefficients are $a,b,c,d$, and $e$. Magnitudes from Cat2 are then transformed into Cat1 magnitudes and the iteration block of the process begins.\\

In each iteration, Cat1 and Cat2 are newly matched with stronger restrictions: sky separation $< 0.65$; $|B_{\rm Cat1} -B_{\rm Cat2}| <  3[\langle \sigma(B_{\rm Cat1}) + \langle \sigma(B_{\rm Cat2}) \rangle]$; $|V_{\rm Cat1} -V_{\rm Cat2}| <  3[\langle \sigma(V_{\rm Cat1}) \rangle+ \langle \sigma(V_{\rm Cat2}) \rangle]$, rejecting all stars not fulfilling these requirements. After matching both catalogues, a new second-order fitting is performed:\\

\begin{math}
   B_{\rm Cat1} = a + b {\rm RA}_{\rm Cat2} + c {\rm Dec}_{\rm Cat2} + d B_{\rm Cat2} + e (B_{\rm Cat2}-V_{\rm Cat2})+ f B_{\rm Cat2}^2 + g (B_{\rm Cat2}-V_{\rm Cat2})^2
\end{math}\\

\begin{math}
   V_{\rm Cat1} = a + b {\rm RA}_{\rm Cat2} + c {\rm Dec}_{\rm Cat2} + d V_{\rm Cat2} + e (B_{\rm Cat2}-V_{\rm Cat2})+ f V_{\rm Cat2}^2 + g (B_{\rm Cat2}-V_{\rm Cat2})^2
\end{math}\\

\begin{figure}
\begin{center}
\includegraphics[scale=0.75]{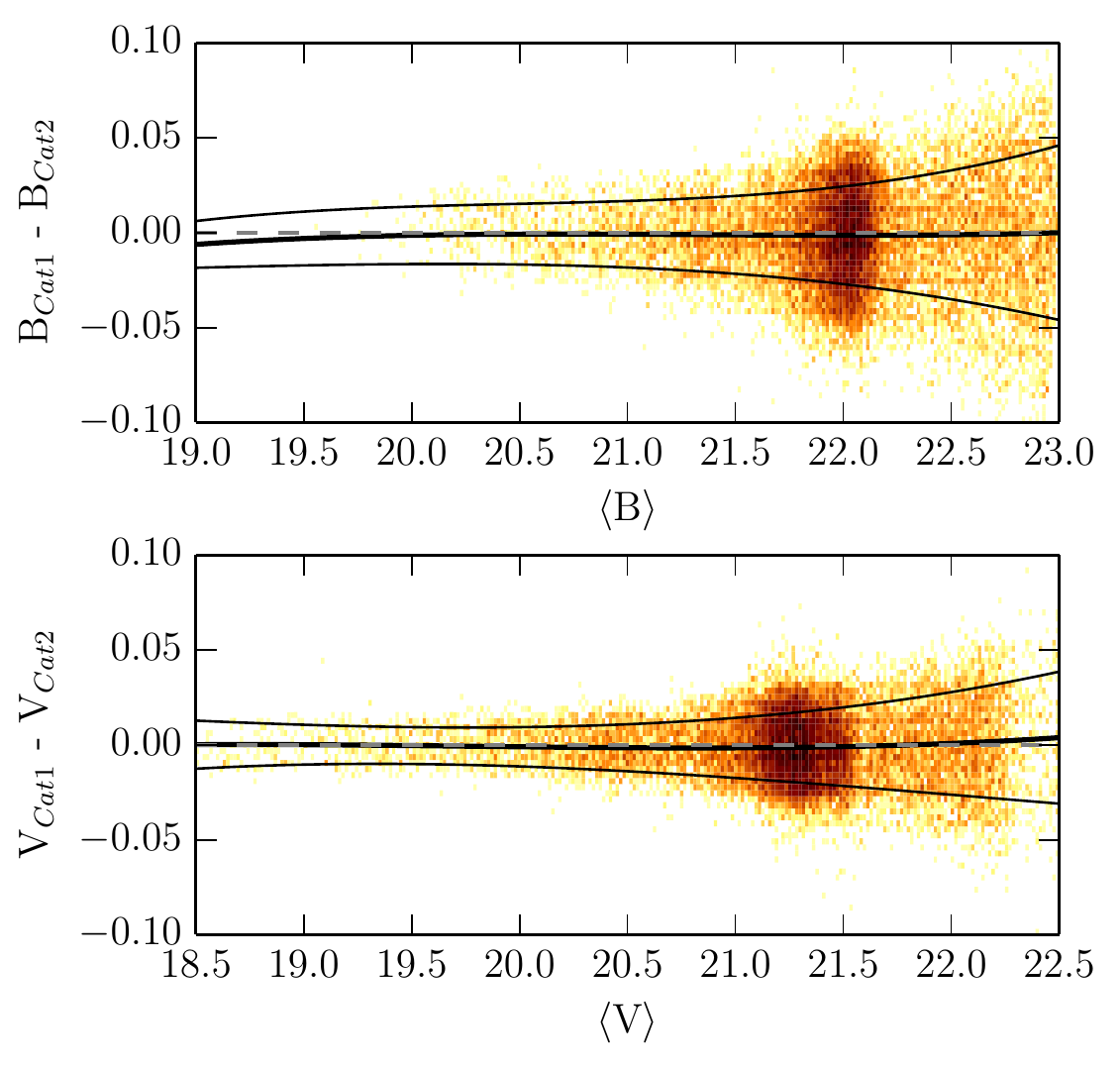}
\caption[Internal calibration between catalogues]{Internal calibration between catalogues. Magnitude differences between Cat1 and Cat2 after calibration. Thick lines represent the average difference between magnitudes, while thin lines show the $\pm1\sigma$ standard deviation.}
\label{fig:Calibration}
\end{center}
\end{figure}

Being $p_i = (a_i, b_i, c_i, d_i, e_i, f_i, g_i)$ the fitting parameters in the $i^{th}$ iteration, the whole process converges when $|p_{i} -p_{i-1}| \leq p_{i}\times10^{-4}$. Each fitting is performed applying $\sigma$-weights in each iteration. A total of 26179 stars were used in the calibration, which ended with $\chi_{\rm red}^B = 1.02$ and $\rm \chi_{\rm red}^V = 1.04$ for the filters $B$, and $V$ respectively. In Fig.~\ref{fig:Calibration}, we show the resulting differences between both catalogues after the calibration. Results show a very good consistency with an internal dispersion below 0.05 mag for all the magnitude range.\\

\begin{figure}
\begin{center}
\includegraphics[scale=0.75]{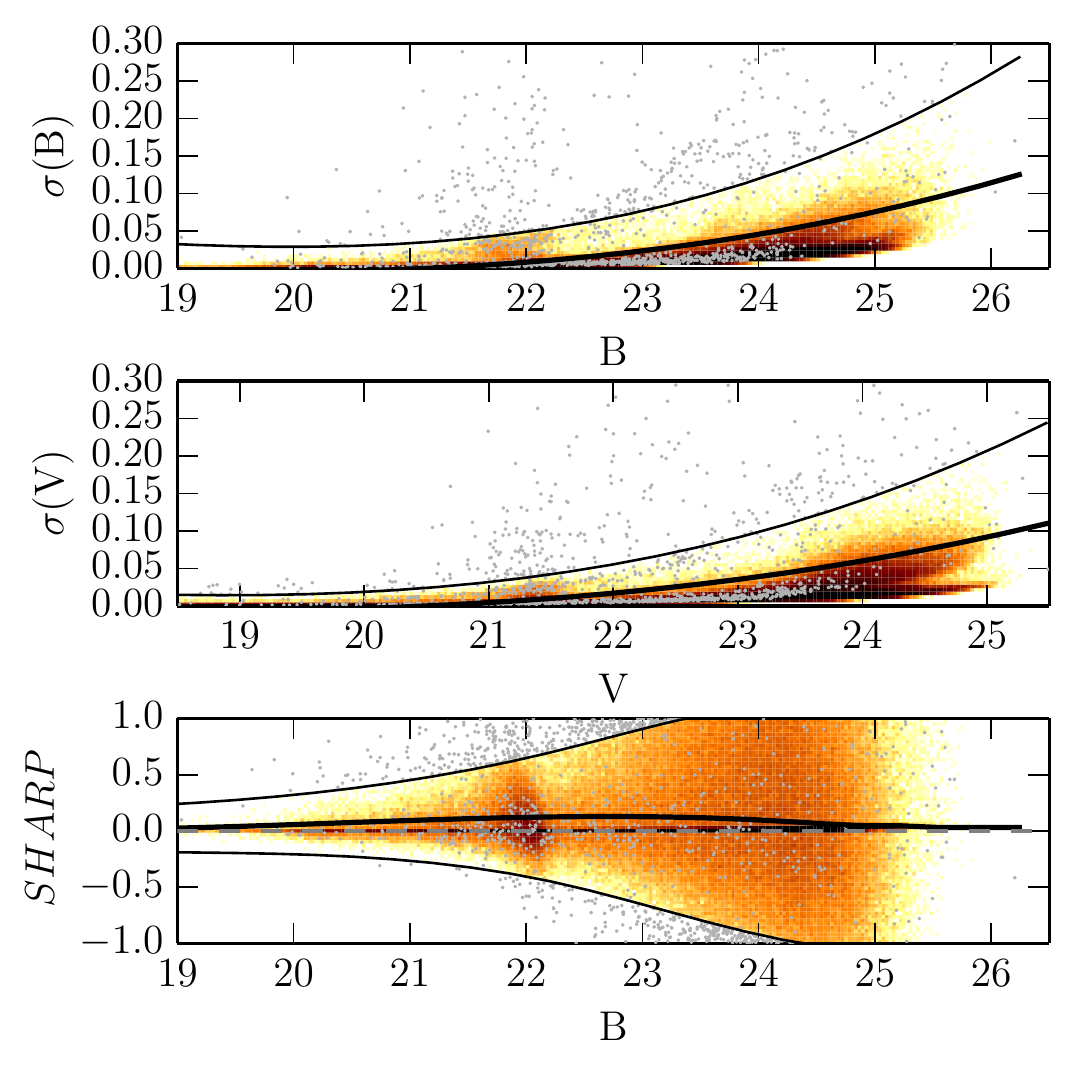}
\caption[Cleaning criteria adopted]{Criteria adopted for cleaning the photometric list from non-stellar objects. From top to bottom, $\sigma(\rm B)$, $\sigma(\rm V)$, and \textit{SHARP} versus magnitude of the measured stars. The thin lines show the $\sigma$ rejection limit adopted to clean the photometric list, $3\sigma$ for magnitude error, and $2\sigma$ for the \textit{SHARP} parameter. Grey dots show the rejected stars. Only 10 per cent of the rejected stars is shown.}
\label{fig:Errors_cut2}
\end{center}
\end{figure}

Both photometries were corrected of distance modulus and reddening. The reddening was corrected using dust maps by \citet*{Schlafly2011} for both filters $(B,V)$. No reddening gradients were observed within the photometry boundaries in these maps, being $E(B-V) = 0.019$ mag the adopted reddening. Finally, we cleaned our photometry of likely non-stellar objects and stars with high magnitude errors on the basis of $\sigma$, $\chi^2$ and \textit{SHARP} parameters provided by \textsc{Daophot} for each star in Cat1, and $\sigma$, and \textit{q} provided by \textsc{Dophot} in Cat2. Fig.~\ref{fig:Errors_cut2} shows the rejection criteria followed for Cat1.\\

\subsection{Unicity}\label{Ap:Internal_Calibration:Data_Unicity}

In order to avoid repeated stars, we performed a last matching between Cat1 and the calibrated Cat2. The whole calibrated set of stars was used, including those with slightly worse measurements, and not isolated stars. More permissive photometric differences between matched stars were allowed in this procedure, adopting a $(3\sigma+0.25 \rm Mag)$ as the maximum permitted magnitude difference. The final photometric list was obtained as the compilation of all stars from Cat1 and those present in Cat2 which are not in Cat1.\\

\subsection{Completeness}\label{Ap:Internal_Calibration:Data_Completeness}

A detailed analysis of the stellar populations would require meticulous crowding tests. These are normally based on the injection of millions of artificial stars in the original CCD images, retrieving the completeness as a function of magnitude, colour, and CCD position. This process requires, however, the considerable effort of re-obtaining the photometry. Since such detailed analysis is beyond the scope of this work, we did not carry out these tests. Notwithstanding, some information about the completeness of the final sample is needed to perform our analysis.\\

To inquiry the completeness of the final catalogue without the use of artificial star tests, we made use of the completeness derived by \citet{deBoer2012} for Cat2 at different concentric elliptical regions. These were defined within ellipses centred in Fornax with radius of 0.1, 0.2, 0.3, and 0.4 deg measured over their semi-major axes ($r_{\rm ell}$). For stars within an elliptical region, we defined $C_1({\rm Mag})$ and $N_1({\rm Mag})$ as the completeness and the number of stars at a given magnitude for Cat1, respectively. We use the same definition for Cat2 catalogue: $C_2({\rm Mag})$ and $N_2({\rm Mag})$. These definitions allow us to derive the completeness of Cat1 as a function of magnitude as
$$
   C_1({\rm Mag}){\rm d Mag} = C_2({\rm Mag}) \frac{N_1({\rm Mag})}{N_2({\rm Mag})}{\rm d Mag}
$$

By defining $N_{1|2}({\rm Mag})$ as the number of stars at a given magnitude present in Cat1 which are not in Cat2, the completeness of the final photometric list for a specific elliptical region can be expressed as
$$
   C_T({\rm Mag}){\rm d Mag} = C_2({\rm Mag}) \frac{N_2({\rm Mag})+N_{1|2}({\rm Mag})}{N_2({\rm Mag})}{\rm d Mag}
$$

\begin{figure}
\begin{center}
\includegraphics[scale=0.75]{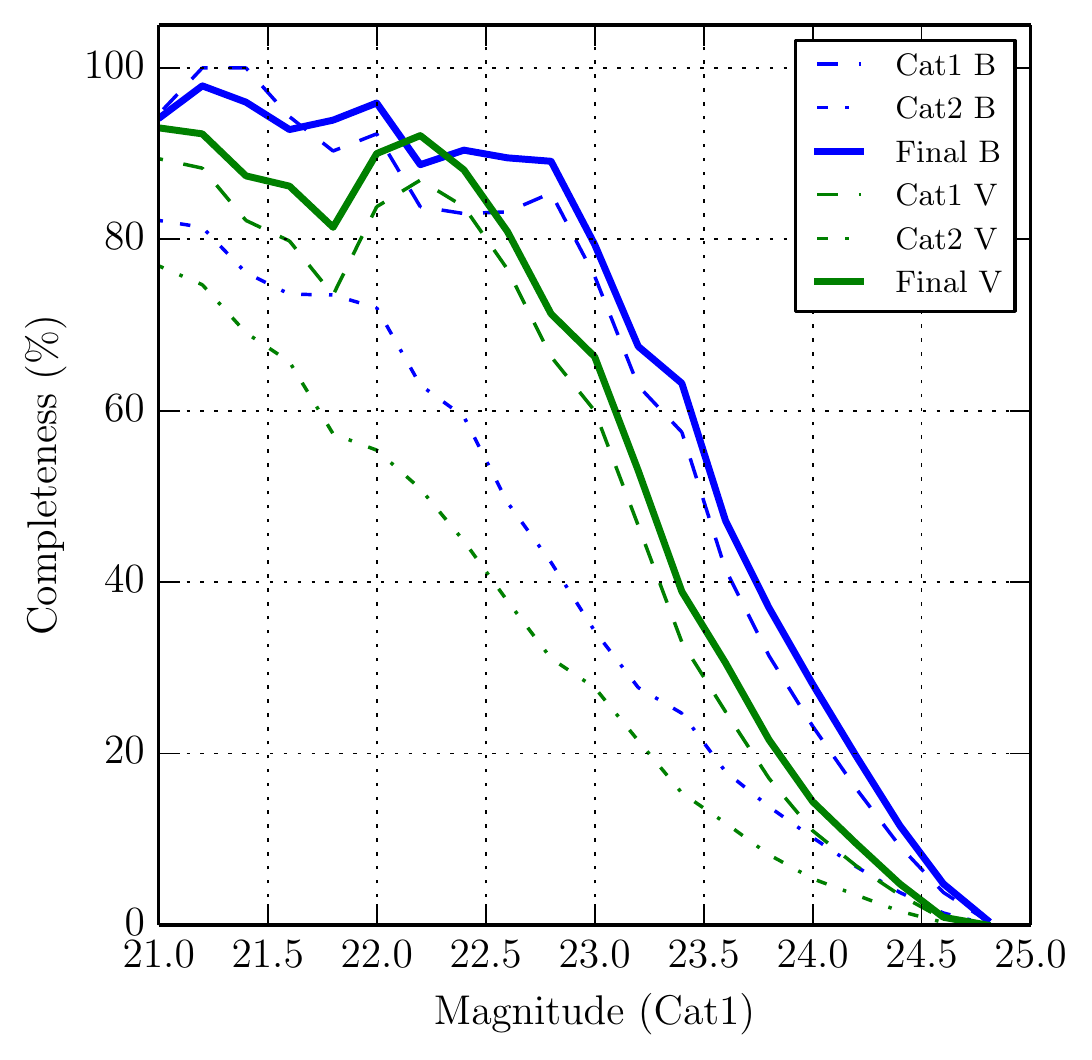}
\caption[Completeness in the most over-crowded regions]{Completeness as a function of magnitude for both filters within an ellipse of $r_{\rm ell} = 0.1^\circ$ \citep{deBoer2012}. Dashed and dot-dashed lines represent the completeness for Cat1 and Cat2, respectively.}
\label{fig:Completeness}
\end{center}
\end{figure}

Completeness is affected by crowding and can drop substantially in the innermost regions of the galaxy, where the stellar density is higher. This may impact our results, especially when deriving radial density profiles of the stellar populations. We calculated the completeness for both filters as a function of magnitude for the central region defined by \citet{deBoer2012}. Results are shown in Fig.~\ref{fig:Completeness}. We decided to keep $B = 23.00$ and $V = 22.67$ as our deepest magnitudes for all further analysis, a completeness level of at least 80 per cent being assured in all the cases.\\

\subsubsection{Spatial correction for completeness}\label{Ap:Internal_Calibration:Data_Completeness:Spatial_Correction}

Completeness drops significantly towards the centre of the galaxy. In order to properly correct this effect, coordinate-dependent crowding test would be required. However, we only have information about the average completeness within the elliptical regions defined by \citet{deBoer2012}. Assuming an axisymmetric completeness distribution, we interpolated the completeness, $C_T({\rm Mag})$, as a function of the elliptical radius ($r_{\rm ell}$) and magnitude. The interpolant was determined by triangulating the completeness, and constructing a piecewise cubic interpolating Bezier polynomial on each triangle using a Clough--Tocher scheme \citep[see, for example,][]{Clough_Toucher}. The interpolated surface was then extrapolated to the actual limits of the photometry in terms of $r_{\rm ell}$ and magnitude. Two completeness functions were created, one per filter. The associated completeness to each star in the photometry was chosen to be the lowest value obtained through these functions, by using the star coordinates $C_T(r_{\rm ell}, B)$ and $C_T(r_{\rm ell}, V)$. This method allows us to obtain the completeness of our photometry as a function of $r_{\rm ell}$, $B$, and $V$ magnitudes.\\

\subsection{Background removal}\label{Ap:Internal_Calibration:Cleanning_Background}

Background level estimate is another relevant problem in examining density profiles of low-surface-brightness galaxies. Overestimating the background leads to artificially truncated profiles, while underestimating it produces abnormally shallow ones with potentially non-convergent integral light. Our sample covered area allows us to infer the effect of this 'background' consisting in foreground stars and background galaxies. This was extracted from the sources detected within each bundle (see Section~\ref{Cap:CMD:CMDsample_1}) and outside the tidal radius of Fornax, given by \citet{Battaglia2006}, $r_{\rm tidal} = 69.4^\prime$. Despite extending over almost the full Fornax CMD, they gather mainly around $\rm M_{V} = 3.75$ with colours $(B-V)_0$ spanning from 0 to 1. No obscuration has been reported within the galaxy, being its internal absorption negligible \citep{Lisenfeld1998, Bouchard2006}. Therefore, we can assume a uniform level of contamination along the whole body of the galaxy. This background contribution was removed from the final radial density profiles obtained in Section~\ref{Cap:Radial_Profiles:Fitting_profiles}.\\

\section{Stellar radial density profiles}\label{Ap:Radial_Density_Profiles_Tables}

Resulting stellar radial density profiles are listed for each subpopulation in Table~\ref{tab:Radial_Profiles_annul1_Subpop}, while Table~\ref{tab:Radial_Profiles_annul1_Total} shows the same profiles for the total population.\\
 
\begin{itemize}
\item Column 1: subpopulation;
\item Column 2: average radius in pc of the stars within the elliptical region;
\item Column 3: stellar surface density corrected from completeness and contaminants.
\end{itemize}

\begin{table}
  \caption{Stellar densities as a function of galactocentric radius for each subpopulation using free pattern elliptical regions.}
  \label{tab:Radial_Profiles_annul1_Subpop}
  \begin{tabular}{@{}lcccc}
    \hline
    \hline
    Subpop. & \multicolumn{2}{c}{$\rm r$} & \multicolumn{2}{c}{$I$}\\
            & $\rm [^\prime]$ & $\rm [pc]$ & $\rm[pc^{-2}]$ & $\rm[(^\prime)^{-2}]$\\
    \hline
    HB
    & 4.31  & 170.5 &$(1.48\pm0.08)\times10^{-3}$  & $2.3\pm0.1$ \\
    & 8.37  & 330.9 &$(1.12\pm0.07)\times10^{-3}$  & $1.8\pm0.1$ \\
    & 11.28 & 446.2 &$(1.08\pm0.06)\times10^{-3}$  & $1.7\pm0.1$ \\
    & 14.20 & 561.7 &$(7.4\pm0.5)\times10^{-4}$  & $1.16\pm0.07$ \\
    & 17.19 & 680.1 &$(6.2\pm0.4)\times10^{-4}$  & $0.97\pm0.06$ \\
    & 20.33 & 804.3 &$(4.9\pm0.3)\times10^{-4}$  & $0.76\pm0.05$ \\
    & 24.15 & 955.5 &$(3.7\pm0.2)\times10^{-4}$  & $0.57\pm0.03$ \\
    & 28.64 & 1133.0& $(2.7\pm0.2)\times10^{-4}$  & $0.42\pm0.03$ \\
    & 33.81 & 1337.8& $(1.4\pm0.1)\times10^{-4}$  & $0.22\pm0.02$ \\
    & 43.46 & 1719.2& $(5.2\pm0.8)\times10^{-5}$  & $0.08\pm0.01$ \\
    \hline
    RGB
    & 3.77 & 149.1   &$(8.2\pm0.2)\times10^{-3}$  & $12.8\pm0.4$ \\
    & 6.60 & 261.0   &$(6.2\pm0.2)\times10^{-3}$  & $9.6\pm0.3$ \\
    & 8.79 & 347.9   &$(5.6\pm0.2)\times10^{-3}$  & $8.7\pm0.3$ \\
    & 10.84 & 429.0  &$(4.4\pm0.1)\times10^{-3}$  & $6.9\pm0.2$ \\
    & 12.92 & 510.9  &$(3.5\pm0.1)\times10^{-3}$  & $5.5\pm0.2$ \\
    & 15.11 & 597.6  &$(2.80\pm0.09)\times10^{-3}$  & $4.4\pm0.1$ \\
    & 17.60 & 696.2  &$(1.89\pm0.07)\times10^{-3}$  & $3.0\pm0.1$ \\
    & 20.58 & 814.3  &$(1.30\pm0.05)\times10^{-3}$  & $2.04\pm0.07$ \\
    & 24.84 & 982.6  &$(7.6\pm0.3)\times10^{-4}$  & $1.19\pm0.04$ \\
    & 34.36 & 1359.3 &$(1.23\pm0.08)\times10^{-4}$  & $0.19\pm0.01$ \\
    \hline
    SG
    & 2.65 & 104.8 &$(4.0\pm0.3)\times10^{-3}$  & $6.2\pm0.4$ \\
    & 4.45 & 176.0 &$(2.5\pm0.2)\times10^{-3}$  & $3.9\pm0.3$ \\
    & 6.02 & 238.2 &$(2.4\pm0.2)\times10^{-3}$  & $3.7\pm0.3$ \\
    & 7.41 & 293.0 &$(1.5\pm0.1)\times10^{-3}$  & $2.3\pm0.2$ \\
    & 8.94 & 353.6 &$(9.6\pm0.9)\times10^{-4}$  & $1.5\pm0.1$ \\
    & 10.24 & 405.0& $(5.9\pm0.7)\times10^{-4}$  & $0.9\pm0.1$ \\
    & 11.82 & 467.5& $(4.8\pm0.5)\times10^{-4}$  & $0.75\pm0.08$ \\
    & 14.13 & 559.2& $(3.5\pm0.4)\times10^{-4}$  & $0.54\pm0.06$ \\
    & 16.59 & 656.4& $(1.1\pm0.2)\times10^{-4}$  & $0.17\pm0.03$ \\
    & 32.61 & 1290.2& $(2\pm1)\times10^{-5}$  & $0.03\pm0.02$ \\
    \hline
    MS
    & 2.63 & 104.0  &$(1.93\pm0.05)\times10^{-2}$  & $30.2\pm0.8$ \\
    & 4.63 & 183.0  &$(1.61\pm0.05)\times10^{-2}$  & $25.3\pm0.7$ \\
    & 6.12 & 241.9  &$(1.12\pm0.04)\times10^{-2}$  & $17.5\pm0.6$ \\
    & 7.48 & 296.0  &$(7.3\pm0.3)\times10^{-3}$  & $11.5\pm0.4$ \\
    & 8.83 & 349.4  &$(5.7\pm0.2)\times10^{-3}$  & $8.9\pm0.3$ \\
    & 10.25 & 405.6 &$(3.6\pm0.2)\times10^{-3}$  & $5.6\pm0.2$ \\
    & 11.78 & 466.1 &$(2.4\pm0.1)\times10^{-3}$  & $3.7\pm0.2$ \\
    & 13.84 & 547.6 &$(1.74\pm0.08)\times10^{-3}$  & $2.7\pm0.1$ \\
    & 16.71 & 661.2 &$(7.9\pm0.4)\times10^{-4}$  & $1.24\pm0.06$ \\
    & 27.49 & 1087.4& $(6\pm1)\times10^{-5}$  & $0.09\pm0.02$ \\
    \hline
    YMS
    & 3.65 & 144.4  &$(9\pm1)\times10^{-4}$  & $1.4\pm0.2$ \\
    & 5.31 & 209.9  &$(6.5\pm0.8)\times10^{-4}$  & $1.0\pm0.1$ \\
    & 8.16 & 322.7  &$(5.2\pm0.6)\times10^{-4}$  & $0.8\pm0.1$ \\
    & 9.64 & 381.4  &$(4.3\pm0.6)\times10^{-4}$  & $0.67\pm0.09$ \\
    & 11.36 & 449.3 &$(3.0\pm0.5)\times10^{-4}$  & $0.46\pm0.07$ \\
    & 11.66 & 461.1 &$(1.2\pm0.3)\times10^{-4}$  & $0.19\pm0.04$ \\
    & 11.70 & 462.9 &$(7\pm2)\times10^{-5}$  & $0.12\pm0.03$ \\
    & 13.66 & 540.4 &$(5\pm1)\times10^{-5}$  & $0.08\pm0.02$ \\
    & 17.99 & 711.6 &$(3.7\pm0.9)\times10^{-5}$  & $0.06\pm0.01$ \\
    & 33.14 & 1310.9& $(5\pm4)\times10^{-6}$  & $0.007\pm0.006$ \\
    \hline
  \end{tabular}
\end{table}

\begin{table}
  \caption{Stellar densities as a function of galactocentric radius for each subpopulation using free pattern elliptical regions.}
  \label{tab:Radial_Profiles_annul1_Total}
  \begin{tabular}{@{}lcccc}
    \hline
    \hline
    Subpop. & \multicolumn{2}{c}{$\rm r$} & \multicolumn{2}{c}{I}\\
            & $\rm [^\prime]$ & $\rm [pc]$ & $\rm[pc^{-2}]$ & $\rm[(^\prime)^{-2}]$\\
    \hline
    Total
    & 3.09 & 122.3  &$(8.95\pm0.09)\times10^{-2}$  & $140\pm1$ \\
    & 5.77 & 228.2  &$(6.92\pm0.08)\times10^{-2}$  & $108\pm1$ \\
    & 7.81 & 309.1  &$(5.26\pm0.06)\times10^{-2}$  & $82.3\pm0.9$ \\
    & 9.74 & 385.1  &$(4.11\pm0.05)\times10^{-2}$  & $64.3\pm0.8$ \\
    & 11.74 & 464.5 &$(3.24\pm0.04)\times10^{-2}$  & $50.7\pm0.6$ \\
    & 13.92 & 550.8 &$(2.52\pm0.03)\times10^{-2}$  & $39.4\pm0.5$ \\
    & 16.33 & 646.1 &$(1.86\pm0.02)\times10^{-2}$  & $29.2\pm0.3$ \\
    & 19.53 & 772.8 &$(1.26\pm0.01)\times10^{-2}$  & $19.8\pm0.2$ \\
    & 24.06 & 951.9 &$(6.87\pm0.08)\times10^{-3}$  & $10.8\pm0.1$ \\
    & 34.68 & 1371.9& $(1.31\pm0.04)\times10^{-3}$  & $2.06\pm0.06$ \\
    \hline
  \end{tabular}
\end{table}

\label{lastpage}
\end{document}